\documentclass[pra,twocolumn,preprintnumbers,amsmath,amssymb,superscriptaddress,showpacs]{revtex4-1}

\usepackage{amsfonts}
\usepackage{amssymb}
\usepackage[dvips]{graphicx}
\usepackage{color}
\usepackage{amsmath}
\usepackage{blkarray}

\newcommand{\ket}[1]{\left|#1\right>}

\begin{document}

\title{The dynamics of two entangled qubits exposed to classical noise: role of spatial and temporal noise correlations}
\author{Piotr Sza\'{n}kowski}
\author{Marek Trippenbach}
\affiliation{
Institute of Theoretical Physics, University of Warsaw, ul. Pasteura 5, PL 02-093 Warszawa, Poland}
\author{\L{}ukasz Cywi\'{n}ski}
\affiliation{Institute of Physics, Polish Academy of Sciences, al.~Lotnik{\'o}w 32/46, PL 02-668 Warszawa, Poland}
\author{Y. B. Band}
\affiliation{
Department of Chemistry, Department of Physics,
Department of Electro-Optics, and the Ilse Katz Center for
Nano-Science, Ben-Gurion University, Beer-Sheva 84105, Israel}

\begin{abstract}
We investigate the decay of two-qubit entanglement caused by the influence of classical noise. We consider the whole spectrum of cases ranging from independent to fully correlated noise affecting each qubit. We take into account different spatial symmetries of noises, and the regimes of noise autocorrelation time. The latter can be either much shorter than the characteristic qubit decoherence time (Markovian decoherence), or much longer (approaching the quasi-static bath limit). We express the entanglement of two-qubit states in terms of expectation values of spherical tensor operators which allows for transparent insight into the role of the symmetry of both the two-qubit state and the noise for entanglement dynamics.
\end{abstract}


\maketitle

\section{Introduction}  \label{Sec:Intro}

Quantum entanglement is an information resource \cite{Steane_RPP98,Horodecki_RMP09}. Unfortunately, contact with the environment adversely affects quantum systems: when a small system in a pure quantum state is left in contact with a large environment, the coherence of the system decays \cite{Leggett_RMP87,Weiss,Efrat1,Efrat2,Zurek_RMP03}. Since an entangled state of a multi-partite system is necessarily a superposition state of two or more product states, its entanglement is extremely fragile when exposed to environmental fluctuations \cite{Aolita_arXiv14}. While, in principle, the disentanglement occurs due to unitary dynamics of the total system consisting of the  initially entangled subsystem of interest and its environment, such an approach requires full knowledge of the environmental Hamiltonian and of the Hamiltonian describing the subsystem-environment coupling. This knowledge is often lacking. Furthermore, even if the full Hamiltonian is known, the exact solution of the problem of the joint unitary dynamics might be simply too hard to obtain. It is thus often assumed that the influence of the environment  on the quantum subsystem can be mapped onto the interaction of the subsystem of interest with a noise source .

In what follows we consider environmental noise experienced by two initially entangled qubits (the subsystem of interest here).  The noise dominating the decoherence of the qubits can be effectively classical or quantum, depending on the relation between the bath temperature and the relevant energy scale of the subsystem. When considering the process of relaxation (i.e., energy exchange between the qubits and the bath) the energy scale of the subsystem is the qubit energy splitting, $\hbar\Omega$, and in order for the noise to be classical, one needs $\hbar\Omega <  k_{B}T$, where $T$ is the temperature of the bath \cite{Schoelkopf_spectrometer}. However, for dephasing of the qubits, the low-frequency fluctuations of the environment have dominant influence, and the corresponding energy scale of interest is $\ll  k_{B}T$ for many types of qubits, including superconducting circuits \cite{Paladino_RMP14}, many kinds of spin qubits in semiconductors \cite{Fischer_SSC09, Cywinski_APPA11}, and ion trap qubits \cite{Monz_PRL11, Schindler_NJP13}. Quantum noise is most often modeled using a bosonic environment \cite{Weiss, Schoelkopf_spectrometer, Makhlin_CP04, Pokrovsky} for which the pure dephasing problem (with the coupling only to the ${\sigma}_{z}$ operator of the qubit) can easily be solved \cite{Duan_PRA98}, but the case of general coupling to an environment which causes decoherence is a challenging problem for environments with nonzero correlation time \cite{Weiss, Burkard_PRB09}. In this paper we assume that the environmental noise is classical, and focus on the influence of its spatial and temporal correlations on the entanglement decay of two qubits.

Much work has been reported on the entanglement dynamics of two qubits in the presence of classical environmental noise. The case of pure dephasing has received considerable attention.
Calculations for white noise (i.e., noise having negligible temporal correlations, leading to Markovian decoherence) were carried out both in the case of spatially uncorrelated \cite{Yu_PRB03,Yu_OC06,Ann_PRB07} and correlated noise \cite{Yu_OC06} acting on the two qubits.  For fluctuations with finite correlation time (leading to non-Markovian decoherence), pure dephasing due to two mutually spatially uncorrelated \cite{Yu_OC10, Benedetti_PRA13, Zhou_QIP10} and correlated \cite{Ban_OC08, Zhou_QIP10, Bellomo_PRA10, Benedetti_PRA13} noise was considered. The case of general coupling of qubits to noise also received attention; uniaxial transverse coupling to spatially correlated white noise was considered in Ref.~\cite{Corn_QIP09}, and the case of transverse coupling to a more general class of noise (of $1/f$ type, both Gaussian and non-Gaussian) was considered in Refs.~\cite{Benedetti_PRA13,Bellomo_PRA10}, and in \cite{De_PRA11,Brox_JPA12} where the influence qubit-qubit coupling was also studied.

Here we consider the dynamics of two-qubit entanglement when the qubits having both longitudinal and transverse couplings to a classical Gaussian noise field \cite{Budimir_JSP87, Aihara_PRA90, Szankowski_PRE13}.  We treat noise which exhibits correlations on a length scale of the inter-qubit separation, and which has either negligible or finite autocorrelation time. A possible physical realization of this picture is a system of spin qubits \cite{Fischer_SSC09, Cywinski_APPA11, deLange_Science10} interacting with fluctuating magnetic field, or an array of qubits in an ion trap \cite{Monz_PRL11,Schindler_NJP13}. The fluctuations could be caused by an ensemble of paramagnetic spins located in the neighborhood of the qubits (e.g., impurity spins), or they could be generated in a nearby solid-state based magnet. The properties of the noise that we shall assume correspond to such a setting. Noise generated by many weakly correlated sources can be expected to have Gaussian statistics. Due to the long-range nature of magnetic forces, the random field has a finite correlation length, so that, depending on the relation between the inter-qubit distance and this length, the limits of fully correlated and completely uncorrelated noises can be realized. Finally, in the presence of an external magnetic field the total system (and the noise in particular) should exhibit at least cylindrical symmetry with respect to the axis distinguished by the field.

In our analysis we use the language of spherical tensor operators, which allows for transparent discussion of the role of rotational symmetry of quantum states in decoherence and disentanglement processes resulting from coupling to noise. The averaging over noise realizations is carried out by using a second-order cumulant expansion of relevant operator products \cite{Szankowski_PRE13}.  The cumulant expansion approach is exact in the pure dephasing case, but in the presence of transverse noise it is exact only when the noise has a white spectrum \cite{Fox_JMP74,Budimir_JSP87}.  For strong coupling to slow fluctuations, the quasi-static bath limit \cite{Falci_PRL05,Taylor_QIP06} is reached (i.e., the average is over random fields static on the timescale of entanglement decay) and the second-order cumulant expansion for transverse noise fails (in fact a re-summation of all cumulants is needed in such a case \cite{Makhlin_PRL04,Cywinski_PRA14}). We resort to numerical simulations in this limit, and to an approximate analytical theory (applicable when the qubit energy splitting is much larger than the noise amplitude) in which the influence of transverse noise is treated as a second-order correction to the qubit splitting.

In the calculations we focus on the environmental noise induced dynamics of maximally entangled two-qubit Bell states, and we show that their rotational symmetry is the main source of the difference of their sensitivity to correlated noise. For example the spherically symmetric singlet state $\ket{\Psi_-}$ is immune to fully correlated noise, in contradistinction to the other three Bell states. Furthermore, the symmetries of the noise itself also lead to significant differences in the character of entanglement decay: the presence of transverse noise is necessary for the sudden death of entanglement \cite{Zyczkowski_PRA01, Yu_PRL04, Yu_Science09} to occur, and lowering of noise symmetry from a full rotational one (i.e., isotropic noise) to a cylindrical one causes the decay of the entanglement of $\ket{\Psi_{+}}$ Bell state to be different from the decay of entanglement of $\ket{\Phi_{\pm}}$ Bell states. These results have a transparent interpretation when the formalism of spherical tensor operators is used.

The paper is organized in the following way. In Sec.~\ref{Sec:Model} we introduce the Hamiltonian of the two-qubit system and we discuss the statistical properties of the noise field affecting it. In Sec.~\ref{Sec_Th_app} we introduce the spherical tensor operators, which we use to decompose the two-qubit density matrix into parts with distinct transformation properties under spatial rotations. We also show how a measure of entanglement (the commonly used concurrence \cite{Wooters_PRL98}) of a Bell-diagonal mixed state of two qubits can be expressed by averages of products of appropriate spherical tensor operators and spin operators, and we introduce the cumulant expansion for the superoperator governing the evolution of the density matrix of the two qubits.
Then in Sec.~\ref{sec:phase} we present an exact solution for the pure dephasing case, both for white and colored noise, and consider various degrees of correlation between the two noises. In Sec.~\ref{sec:transverse} we extend our analysis to include the transverse noise, and we consider both the isotropic noise and noise with lower symmetry, for the whole spectrum of cases: from independent to fully correlated noises affecting the two qubits. In the case of white noise we obtain an exact analytical solution, while in the case of noise with finite correlation time we present approximate solutions, using the effective pure dephasing Hamiltonian and also the cumulant expansion carried out to the second order. Finally, we compare our analytic results with numerical simulations.  Section~\ref{sec:conclusions} contains the summary and conclusions.

\section{The model}  \label{Sec:Model}

We consider an open system consisting of two entangled qubits which do not interact with each other. A constant magnetic field ${\bf B}_0$ is applied to each qubit and its direction defines the quantization axis (the $z$-axis) and the energy splitting of the qubits, $\Omega = g \mu B_0$, where $g$ is an effective $g$-factor and $\mu$ is the Bohr (or nuclear) magneton (we set $\hbar=1$). The system Hamiltonian is
\begin{equation}\label{ctr_field}
H_S = \Omega ( J^{(1)}_z +J^{(2)}_z ) ,
\end{equation}
where $J^{(n)}_i = \sigma^{(n)}_i/2$ is the $i$th component of spin operator of the $n$th qubit.  We assume that the effective $g$-factors of the two qubits are the same. This assumption is not always true; e.g., for self-assembled quantum dots, the $g$-factors of confined electrons and holes vary from dot to dot. However, as explained below, without this assumption, full correlation between noise experienced by the two qubits is not possible.

Interaction of the qubits with the environment is modeled by coupling them to an external magnetic noises specified by the Larmor frequencies $\boldsymbol{\omega}^{(n)}(t)$ via the stochastic Hamiltonian
\begin{equation}\label{stochastic_model}
H_{SE} =  \boldsymbol{\omega}^{(1)}(t)\cdot\mathbf{J}^{(1)} + \boldsymbol{\omega}^{(2)}(t) \cdot\mathbf{J}^{(2)} .
\end{equation}
As discussed in Sec.~\ref{Sec:Intro}, the noises are assumed to be represented by Gaussian stochastic processes. Furthermore, we assume that they are stationary. Noise with Gaussian statistics are completely specified in terms of their first two moments, the averages and the correlation functions:
\begin{eqnarray}
&&\overline{\boldsymbol{\omega}^{(n)}(t)} = 0\label{general_avg}\\
&&\overline{\omega_i^{(n)}(t)\omega_j^{(m)}(t')} = \kappa^{(n,m)}_{ij}(t-t') .
\label{kappas}
\end{eqnarray}
Here $\overline{(\ldots)}$ denotes stochastic average, $\omega_i^{(n)}$ ($i=x,y,z$ and $n=1,2$) are the orthogonal components of vector processes $\boldsymbol{\omega}^{(n)}$ and, without the loss of generality, we set the averages (\ref{general_avg}) to zero. We assume that $\kappa^{(n,m)}_{ij}  \propto  \delta_{ij}$, i.e., the noise in orthogonal directions are uncorrelated.

The properties of the autocorrelations $\kappa^{(n,n)}_{ii}(t)$, and cross-correlations $\kappa^{(n,m)}_{ii}(t)$ with $n \neq  m$, depend on the nature of the fluctuating magnetic field.  Here we consider the physical picture situation corresponding to the generation of the fluctuating field by many sources:
\begin{equation}
\mathbf{b}(\mathbf{r},t) = \sum_{k} \mathbf{b}_{k}(\mathbf{r}-\mathbf{r}_{k},t) , \label{eq:b}
\end{equation}
where $\mathbf{r}_{k}$ are the locations of field sources, and $\mathbf{b}_{k}(\mathbf{r},t)$ are the fluctuating fields generated by these sources.  Even if $\mathbf{b}_{k}(\mathbf{r},t)$ are represented by stationary, but not necessarily Gaussian stochastic processes, the Gaussian statistics of the total field $\mathbf{b}(\mathbf{r},t)$ emerges due to the central limit theorem \cite{Ash_08} when the number of weakly correlated noise sources is large.  Then, the noise affecting  the two qubits is proportional to $\mathbf b(\mathbf{r},t)$ at the locations $\mathbf{R}_{n}$ of the qubits:
\begin{equation}
\boldsymbol{\omega}^{(n)}(t) \propto g \, \mathbf{b}(\mathbf{R}_{n},t) ,
\end{equation}
We assumed that the proportionality factor is the same for both qubits (i.e., their $g$-factors are the same).

In what follows, we neglect any effects of finite propagation time in Eq.~(\ref{eq:b}), which allows us to derive the relation
\begin{equation}
\kappa^{(1,2)}_{ii}(t-t') = \kappa^{(2,1)}_{ii}(t-t') \equiv \kappa^{\times}_{ii}(t-t') .
\end{equation}
Furthermore the argument of the above functions is in fact $|t-t'|$, just like in the case of the autocorrelation $\kappa^{(n,n)}_{ii}(t-t')$. Note that this is not an obvious property, since in the general case (with finite speed of signal propagation), only $\kappa^{(1,2)}(t-t') = \kappa^{(2,1)}(t'-t)$ is guaranteed.

The $i$th component of the noises are thus characterized by three correlation functions $\kappa^{(n,n)}_{ii}(t)$, $i = 1,2,3$, describing the temporal correlations of $\omega_{i}^{(n)}$ process experienced by qubit $n = 1$ and $2$, and the $\kappa^{\times}_{ii}(t)$ describing the cross-correlation between the two processes. In the general case of $\mathbf{b}(\mathbf{r},t)$ with finite correlation time and correlation length, we have to deal with three independent functions.  Using the above physical model of the noise source, it is clear that for spatially uniform $\mathbf{b}(\mathbf{r},t) = \mathbf{b}(t)$ field, i.e., when the $\mathbf b(\mathbf{r},t)$ field is uniform on a length scale larger than the inter-qubit spacing, the correlation between the $\boldsymbol{\omega}^{(1)}(t)$ and $\boldsymbol{\omega}^{(2)}(t)$ is perfect and we have $\kappa^{(1,1)}_{ii}(t) = \kappa^{(2,2)}_{ii}(t) = \kappa^{\times}_{ii}(t)$.  This we take as the definition of full correlation of the noise felt by the two qubits.  In the opposite limit of independent noises, $\kappa^{\times}_{ii}(t) = 0$, and the noise is described by the two remaining autocorrelation functions.

The correlations of $\boldsymbol{\omega}^{(1)}(t)$ and $\boldsymbol{\omega}^{(2)}(t)$ can be easily described for the case when $\mathbf{b}(\mathbf{r},t)$ has autocorrelation time much smaller than any other timescale of the system, i.e., when the autocorrelation time is much smaller than the qubit population and coherence decay timescales, $T_{1}$ and $T_{2}$. This is the white noise limit that leads to Markovian decoherence. If we further assume that the squared amplitude of the noise is the same for the two qubits, we have
\begin{align}
\kappa^{(n,n)}_{ii}(t-t') {}& \approx \frac{2}{T_{ii}} \delta(t-t') , \label{eq:knn_white}  \\
\kappa^{\times}_{ii}(t-t') {}& \approx \gamma_{i} \frac{2}{T_{ii}} \delta(t-t') ,  \label{eq:kx_white} \end{align}
where $\gamma_{i} \in [0,1]$ parametrize the strength of the correlation between the noise, and $T_{ii}$ characterize the noises along the $i$ axis.


\section{Theoretical Approach} \label{Sec_Th_app}

\subsection{Spherical Tensor Operators} \label{sec:tensors}

In order to describe the symmetry of a state under rotations, its density matrix may be decomposed into parts which transform independently under rotations. For example, the density matrix of a single qubit, which is equivalent to spin $1/2$ system, can be written as a combination of spin operators ${\bf J} = {\boldsymbol{\sigma}}/2$ and the unit matrix $ \mathbf 1$:
\begin{equation}\label{qubit_rho}
\varrho_{\mathrm{qubit}} = \frac 12 \mathbf 1 + 2\left\langle\mathbf{J}\right\rangle\cdot\mathbf{J},
\end{equation}
where  $\langle {\bf J} \rangle = \mathrm{Tr}\left( {\bf J} \varrho_{\mathrm{qubit}}\right)$.
An arbitrary rotation transforms the spin operator into a linear combination of spin operators, while the unit matrix is unaffected by rotations. Hence, (\ref{qubit_rho}) is the desired decomposition, where the part spanned by  ${\bf J}$ is independent of the part spanned by $\mathbf 1$. Although the decomposition (\ref{qubit_rho}) might seem to trivially result from the properties of Pauli matrices $\sigma_i$, in fact this decomposition is a consequence of a more general principle. In the language of group theory, the spin operators are vectorial objects. More precisely, they are proportional to {\em spherical tensor operators} \cite{Sakurai_podrecznik_do_QM} with angular momentum $l=1$ and magnetic number $m$ taking on values of $-1$, $0$ and $1$:
\begin{align}
&T_{l=1,m=0}=J_z, \label{def_10_tensor} \\
&T_{l=1,m=\pm 1} = \mp \frac 1{\sqrt{2}}\left(J_x \pm i J_y\right)\equiv \mp \frac 1{\sqrt 2} J_\pm .\label{def_11_tensor}
\end{align}
The spherical tensor operators $T_{lm}$ are an operator analog of the angular momentum eigenstates $\left|l,m\right\rangle$. For example, one way of defining the spherical tensor with quantum numbers $l$ and $m$ is to require that it satisfy a following set of equations:
\begin{align}
&\left[ J_z , T_{lm}\right] = m \, T_{lm}\label{def_tensor_m} , \\
&\left[ J_\pm, T_{lm}\right] = \sqrt{\left(l\pm m\right)\left(l\pm m + 1\right)} \, T_{l m\pm 1} .\label{def_tensor_l}
\end{align}
These equations are in direct correspondence with a similar set of equations satisfied by the angular momentum eigenstates: $J_z\left| l, m\right\rangle = m\left| l, m\right\rangle$ and $J_\pm\left| l,m\right\rangle = \sqrt{(l\pm m)(l\pm m +1)}\left| l,m\pm 1\right\rangle$. Since the action of an arbitrary rotation on an operator is determined by commutators such as (\ref{def_tensor_m}) and (\ref{def_tensor_l}), it follows that the transformation properties of spherical tensors and angular momentum eigenstates are the same. One of the consequences is that any rotation transforms spherical tensors with angular momentum $l$ into a combination of spherical tensors with the same $l$.

The set of spherical tensor operators forms an orthogonal basis with respect to the scalar product $( A | B ) \equiv \mathrm{Tr}( A^\dagger B)$ in the space of density matrices of spin systems. In particular, the density matrix $\varrho_{s}$ of a spin $s$ system, operating in $2s+1$ dimensional Hilbert space, can be decomposed into linear combination of spherical tensor operators with angular momenta ranging from $l=0$ to $l=2s$ and the unit matrix:
\begin{equation}
\varrho_s = \frac 1{2s+1}\mathbf 1+\sum_{l=0}^{2s}\sum_{m=-l}^l \varrho_{lm}T_{lm},
\end{equation}
where the coefficients
\begin{equation}
\varrho_{lm}=\frac 1{\mathrm{Tr}(T^\dagger_{lm}T_{lm})}\mathrm{Tr}\left( T^\dagger_{lm} \varrho_s\right) ,
\end{equation}
are the expectation values of the spherical tensor operators. The parts of $\varrho_s$ spanned by spherical tensors with fixed $l$ have different symmetry and they transform independently. Therefore, decomposition such as this provides a proper quantification of symmetry properties of a state described by density matrix $\varrho_s$.

The density matrix of a composite system, such as a two qubit state, is spanned by outer products of spherical tensor operators of the constituents. In general, a tensor product operator $T^{(1)}_{l_1 m_1}\otimes T^{(2)}_{l_2 m_2}$ can be decomposed into parts that transform independently with respect to global rotations generated by the total spin operators $J_i = J_i^{(1)}+J_i^{(2)}$, in a fashion similar to the outer product of angular momentum eigenstates. This decomposition is subject to the same rules of addition of angular momenta as products of eigenstates, i.e.,
\begin{equation}\label{ang_mom_addition}
T_{L M(l_1 l_2)}=\sum_{m_1,m_2}\left\langle l_1 ,m_1 ; l_2, m_2 \right|\left. \!L,M\right\rangle T^{(1)}_{l_1 m_1}\!\!\otimes\! T^{(2)}_{l_2 m_2} \,\, ,
\end{equation}
where $\left\langle l_1 ,m_1 ; l_2, m_2 \right|\left. L,M\right\rangle$ are the Clebsch-Gordan coefficients.

\subsection{Decomposition of Bell States}

As a first application we present the decomposition of density matrices of maximally entangled Bell states in the basis of spherical tensor operators. The states are defined as
\begin{align}
&|\Psi_-\rangle = \frac{\left|\uparrow\downarrow\right\rangle-\left|\downarrow\uparrow\right\rangle}{\sqrt 2} = | 0,0 \rangle,\label{singlet}\\
&|\Psi_+\rangle = \frac{\left|\uparrow\downarrow\right\rangle+\left|\downarrow\uparrow\right\rangle}{\sqrt 2} = |1, 0\rangle,\label{triplet}\\
&|\Phi_\pm\rangle =\frac{\left|\uparrow\uparrow\right\rangle\pm\left|\downarrow\downarrow\right\rangle}{\sqrt 2} = \frac{|1 , 1\rangle \pm |1,-1\rangle}{\sqrt 2},\label{NOON}
\end{align}
where $\{ \left|\uparrow\downarrow\right\rangle, \left|\downarrow\uparrow\right\rangle, \left|\uparrow\uparrow\right\rangle , \left|\downarrow\downarrow\right\rangle\}$ are the elements of the product basis of spin eigenstates of qubits, and $\{ |s,m_s\rangle\}^{s=0,1}_{m_s=-s,\ldots,s}$ is the basis of total spin. Combining together definitions of one-qubit tensor operators (\ref{def_10_tensor}), (\ref{def_11_tensor}) with equations (\ref{singlet}), (\ref{triplet}) and (\ref{NOON}), the density matrices are decomposed into
\begin{align}
\varrho_{\Psi_-} &= \frac 14 \mathbf 1 - T^{(1)}_{10}\!\otimes\! T^{(2)}_{10} + T^{(1)}_{11}\!\otimes\! T^{(2)}_{1-1}+T^{(1)}_{1-1}\!\otimes\! T^{(2)}_{11}=\nonumber\\
&= \frac 14 \mathbf 1 +\sqrt 3 \,T_{00(11)}\label{singlet_decomp}, \\[.5cm]
\varrho_{\Psi_+} &= \frac 14\mathbf 1 -T^{(1)}_{10}\!\otimes\! T^{(2)}_{10}- T^{(1)}_{11}\!\otimes\! T^{(2)}_{1-1}-T^{(1)}_{1-1}\!\otimes\! T^{(2)}_{11}=\nonumber\\
&=\frac 14 \mathbf 1-\frac 1{\sqrt 3}T_{00(11)}-2\sqrt{\frac 23}\, T_{20(11)} ,\label{triplet_decomp}\\[.5cm]
\varrho_{\Phi_\pm} &=\frac 14\mathbf 1 + T^{(1)}_{10}\!\otimes\! T^{(2)}_{10}\pm \left(T^{(1)}_{11}\!\otimes\! T^{(2)}_{11}+T^{(1)}_{1-1}\!\otimes\! T^{(2)}_{1-1}\right)=\nonumber\\
&=\frac 14 \mathbf 1  -\!\frac 1{\sqrt 3}T_{00(11)}+\!\sqrt{\frac 23}\, T_{20(11)}\!\pm\left(T_{22(11)}\!+T_{2-2(11)}\right) . \label{NOON_decomp}
\end{align}
Note that all four Bell states are spanned by tensor products of the form $T^{(1)}_{1 m_1}\!\otimes\! T^{(2)}_{1 m_2}$ with total magnetic number $M=m_1+m_2$ equal either $0$ or $\pm2$, without partially polarized tensors with $M=\pm 1$. Moreover, an important feature of the representation of the Bell states is that they do not contain products of form $\mathbf 1^{(1)}\!\otimes T^{(2)}_{1 m_2}$ or $T^{(1)}_{1 m_1}\!\otimes\! \mathbf 1^{(2)}$, which describe single-qubit properties of a state which are unrelated to classical or quantum correlations. This is an intuitive result, since Bell states are maximally entangled, so the ``space'' available in the density matrix cannot be wasted on parts that do not contribute to correlations between qubits. Any arbitrary mixture of Bell states, the so-called \emph{Bell-diagonal} states, also possess these properties.  On the other hand, the so-called \emph{X-states} \cite{Yu_QIC07}, often considered in the context of entanglement decay, in general also include parts spanned by the single-qubit products $\mathbf 1^{(1)}\otimes T^{(2)}_{10}$ and $T^{(2)}_{10}\otimes\mathbf 1^{(2)}$.

In what follows we find that a class of two-qubit states spanned exclusively by $T^{(1)}_{1 m_1}\!\otimes T^{(2)}_{1 m_2}$ satisfying the condition $m_1+m_2 = 0, \pm 2$ play a major role in our considerations. For convenience, we shell refer to this class as {\it fully correlated X-states} or $X_{\mathrm{corr}}$--states. Thus, the class of $X_{\mathrm{corr}}$--states include Bell states, Bell-diagonal states and it is contained in the class of $X$--states and also can be classified as {\it states with maximal disordered subsystems} (also known as states with maximally mixed marginals) \cite{Horodecki_PRA96}.

\subsection{Relation between Symmetry of a Two-qubit State and its Entanglement}

In the following investigation we use the {\it concurrence} \cite{Hill_PRL97,Wooters_PRL98}, which is a commonly used measure of entanglement for two-qubit systems that works for pure and mixed states alike. It is defined as
\begin{equation}
\mathcal C = \text{max}\left\{ 0 , r_1 - r_2-r_3-r_4\right\} ,
\end{equation}
where $r_1  \geq  r_2  \geq r_3 \geq r_4$ are the eigenvalues of the matrix
\begin{equation}
\mathcal R(\varrho) = \sqrt{\varrho \, \tau(\varrho)} ,
\end{equation}
with the time reverse superoperator $\tau$ defined by
\begin{equation}
\tau(\varrho) = (\sigma^{(1)}_y\otimes\sigma^{(2)}_y)\varrho^\ast(\sigma^{(1)}_y\otimes\sigma^{(2)}_y) .
\end{equation}
The concurrence ranges from $\mathcal C=0$ for separable states, to $\mathcal C=1$ for maximally entangled states.

Using the above definition of concurrence, it is difficult to identify which properties of a two-qubit state lead to a given amount of entanglement. In fact, concurrence is not expected to have a straightforward physical interpretation considering its origin as a result of maximization of the average entropy of entanglement over all possible pure-state ensembles  $\{ p_k , |\phi_k\rangle\}$ of a mixed state (i.e., $\varrho=\sum_k p_k |\phi_k\rangle\langle\phi_k|$ with $p_k\geqslant 0$ and $\sum_k p_k =1$) \cite{Hill_PRL97, Aolita_arXiv14}. Remarkably, the spherical tensor formalism lets us establish a connection between concurrence of $X_{\mathrm{corr}}$--states and the symmetry of the state.

In Appendix~\ref{app1} we show that the concurrence of an $X_{\mathrm{corr}}$--state $\varrho$ is given by the following expectation values of outer products of spherical tensors:
\begin{align}
\mathcal C = & \, 2\,\mathrm{max}\Big\{ 0\;,  \nonumber\\ &
 2 \big|\big\langle T^{(1)}_{11}\otimes T^{(2)}_{1\,-1}\big\rangle_{\varrho}\big|-\frac14- \big\langle T^{(1)}_{1 0}\otimes T^{(2)}_{1 0} \big\rangle_{\varrho}, \nonumber \\ 
	& 2 \big|\big\langle T^{(1)}_{11}\otimes T^{(2)}_{1\,1}\big\rangle_{\varrho}\big|-\frac14+ \big\langle T^{(1)}_{1 0}\otimes T^{(2)}_{1 0} \big\rangle_{\varrho}\Big\}, \label{bell_diag_conc} 
\end{align}
where $\left\langle A \right\rangle_\varrho=\mathrm{Tr}(A\varrho)$. Since the spherical tensors $T^{(n)}_{1m}$ are proportional to the spin operators, the concurrence is expressed in terms of measurable quantities such as spin projection correlations. Moreover, these quantities are basis independent, in contrast to a commonly used formula for $X$--states where concurrence is characterized by matrix elements of the density matrix written in a product basis \cite{Yu_QIC07} (see Appendix~\ref{app1}).

To illustrate how Eq.~(\ref{bell_diag_conc}) is employed, let us use it to show that Bell states are indeed maximally entangled. Utilizing the correspondence (\ref{def_10_tensor}) and (\ref{def_11_tensor}) between spherical tensor operators and spin operators, we find the explicit orthogonality relations $\mathrm{Tr}( T^{(n)\dagger}_{1 m_1} T^{(n)}_{1 m_2}) = \delta_{m_1 m_2}/2$. Using this, we easily calculate the required expectation values by examining the decompositions of Bell states. Thus, for $|\Psi_-\rangle$ we find
\begin{subequations}\label{eq:S_exp_vals}
\begin{align}
 &\big\langle T^{(1)}_{11}\!\otimes\! T^{(2)}_{1-1}\big\rangle_{\varrho_{\Psi_-}}\!\!\!\!=-\big\langle T^{(1)}_{10}\!\otimes\! T^{(2)}_{10}\big\rangle_{\varrho_{\Psi_-}}\!\!\!\!=\frac 14,\label{subeq:S_exp_val_1-1_00}\\
 &\big\langle T^{(1)}_{11}\!\otimes\! T^{(2)}_{11}\big\rangle_{\varrho_{\Psi_-}}\!\!\!\!=0.\label{subeq:S_exp_val_11}
\end{align}
\end{subequations}
Substituting these values into Eq.~(\ref{bell_diag_conc}), we obtain
\begin{equation}
\mathcal C(\Psi_-) = 2\,\mathrm{max}\Big\{ 0 ,\, 2\times\frac 14-\frac 14+\frac 14 ,\,0 -\frac 14
-\frac 14 \Big\} = 1 .
\end{equation}
 For state $|\Psi_+\rangle$ we have
\begin{subequations}\label{eq:T_exp_vals}
\begin{align}
 &\big\langle T^{(1)}_{11}\!\otimes\! T^{(2)}_{1-1}\big\rangle_{\varrho_{\Psi_+}}\!\!\!\!=\big\langle T^{(1)}_{10}\!\otimes\! T^{(2)}_{10}\big\rangle_{\varrho_{\Psi_+}}\!\!\!\!=-\frac 14,\label{subeq:T_exp_val_1-1_00}\\
 &\big\langle T^{(1)}_{11}\!\otimes\! T^{(2)}_{11}\big\rangle_{\varrho_{\Psi_+}}\!\!\!\!=0\label{subeq:T_exp_val_11}
\end{align}
\end{subequations}
which results in
\begin{equation}
\mathcal C(\Psi_+) = 2\, \mathrm{max}\Big\{ 0,\,2\times\frac 14-\frac 14+\frac 14 ,0 -\frac 14-\frac 14 \Big\} = 1.
\end{equation}
Finally, for states $|\Phi_\pm\rangle$ we have
\begin{subequations}\label{eq:NOON_exp_vals}
\begin{align}
&\big\langle T^{(1)}_{11}\!\otimes\! T^{(2)}_{1-1}\big\rangle_{\varrho_{\Phi_\pm}}\!\!\!\!=0,\label{subeq:NOON_exp_val_1-1}\\
&\big\langle T^{(1)}_{10}\!\otimes\! T^{(2)}_{10}\big\rangle_{\varrho_{\Phi_\pm}}\!\!\!\!=\pm \big\langle T^{(1)}_{11}\!\otimes\! T^{(2)}_{11}\big\rangle_{\varrho_{\Phi_\pm}}\!\!\!\!=\frac 14.\label{subeq:NOON_exp_val_11_00}
\end{align}
\end{subequations}
Thus the concurrence is
\begin{equation}
\mathcal C(\Phi_\pm) =2\,\mathrm{max}\Big\{ 0 ,\, 0-\frac 14-\frac 14 ,\,2\times\frac 14-\frac 14+ \frac 14 \Big\} = 1.
\end{equation}

Equation (\ref{bell_diag_conc}) provides another very important advantage. The expectation values can be computed in the Heisenberg picture, where the spherical tensor operators evolve and the states remains unchanged. As we shall see below, switching to the Heisenberg picture proves to be useful for investigating the role of symmetry.

\subsection{Evolution due to fluctuating fields}

Noise invariably brings an initially pure state $|\Psi\rangle$ to a statistical mixture described by a density matrix:
\begin{equation}
\left|\Psi\right\rangle\!\! \left\langle \Psi\right| \xrightarrow{t} \varrho_\Psi(t).
\end{equation}
In the case of evolution driven by stochastic model such as (\ref{stochastic_model}) this process can be understood in the following way. For every realization of the stochastic process $\boldsymbol\omega_0(t)$, denoted here as a six component vectorial function $\boldsymbol\omega(t) \equiv (\boldsymbol\omega^{(1)}(t), \boldsymbol\omega^{(2)}(t))$ chosen at random with the probability distribution $\mathcal P(\boldsymbol\omega_0)$, the noisy Hamiltonian $H=H_S+H_{SE}$ generates a unitary evolution of the initial state:
\begin{equation}\label{single}
  \varrho_\Psi(t;\boldsymbol\omega_0) = U(t;\boldsymbol\omega_0)\left|\Psi\right\rangle\!\!\left\langle \Psi\right|U^\dagger(t;\boldsymbol\omega_0).
\end{equation}
The unitary evolution operator is given by a time-ordered exponential,
\begin{equation}
  U(t;\boldsymbol\omega_0)=\mathcal T\exp\left[-i\int_0^t dt'\left(H_S+H_{SE}(t')\right)\right] .
\end{equation}
Note that the output state $\varrho_\Psi(t;\boldsymbol\omega_0)$ is still pure. However, since it is unknown which realization governs the evolution, the system must be described by a density matrix $\varrho_\Psi(t)$ that is a mixture of all the possible choices,
\begin{equation}\label{rho_mix}
  \varrho_\Psi(t)= \sum_{\boldsymbol\omega_0}\varrho_\Psi(t;\boldsymbol\omega_0) = \int\mathcal\!\!\mathcal D\boldsymbol\omega_0\,\mathcal P(\boldsymbol\omega_0)\,\varrho_\Psi(t;\boldsymbol\omega_0).
\end{equation}
Here $\int \mathcal D\boldsymbol\omega_0$ is a functional integral over the space of real vectorial functions. On the other hand, the statistical mixture of all possible realizations of the fluctuating field is the same as the definition of the average over process $\boldsymbol\omega$ of the stochastic density matrix
$\varrho_\Psi(t; \boldsymbol\omega)$. Hence, the system density matrix is also given by
\begin{equation}
\varrho_\Psi(t) = \overline{\varrho_\Psi(t;\boldsymbol\omega)}.
\end{equation}

As was shown previously, in order to track the degree of entanglement of the Bell states, we only need to know the evolution of the expectation values of the spherical tensor operators, rather than the whole density matrix. In the case of unitary evolution, one simply needs to switch to the Heisenberg picture. The following calculations shows that the same technique can be used in the case of non-unitary evolution driven by a noisy Hamiltonian \cite{Szankowski_PRA14}:
\begin{align}
&\langle A\rangle_{\varrho_\Psi(t)} = \mathrm{Tr}\left( A\varrho_\Psi(t)\right) = \nonumber\\
&=\int\mathcal\!\!\mathcal D\boldsymbol\omega_0\,\mathcal P(\boldsymbol\omega_0)\mathrm{Tr}\left( A \,U(t;\boldsymbol\omega_0)\left|\Psi\right\rangle \!\!\left\langle \Psi\right|U^\dagger(t;\boldsymbol\omega_0)\right) = \nonumber\\
&=\mathrm{Tr}\left( \int\mathcal\!\!\mathcal D\boldsymbol\omega_0\,\mathcal P(\boldsymbol\omega_0)U^\dagger(t;\boldsymbol\omega_0)A\, U(t;\boldsymbol\omega_0)\left|\Psi\right\rangle\!\! \left\langle \Psi\right|\right) = \nonumber\\
&=\langle \overline{A(t)}\rangle_{\varrho_\Psi(0)}=\left\langle\Psi\right|\overline{A(t)}\left|\Psi\right\rangle .
\label{Heisenberg_pic_derivation}
\end{align}
We proceed to derive the explicit formula for the evolution \emph{superoperator} which transforms the initial operator $A$ into $\overline{A(t)}$. From Eqs.~(\ref{Heisenberg_pic_derivation}) we find that $\overline{A(t)} = \overline{U^\dagger(t;\boldsymbol\omega) A\, U(t;\boldsymbol\omega)}$. Applying the Baker-Campbell-Hausdorff formula followed by the Feynman disentangling theorem \cite{Feynman_PR51} yields a formal expression for the evolution superoperator $\mathcal U$,
\begin{equation}\label{evo_superop}
\overline{A(t)} = e^{i t \mathcal H_S}\overline{\mathcal T \exp\left( i\int_0^t dt' {\mathcal{H}'}_{SE}(t')\right)}\,A\equiv \mathcal U \, A ,
\end{equation}
where the superoperators $\mathcal H_S$ and ${{\mathcal H}'}_{SE}(t)$ are defined as
\begin{align}
&\mathcal H_S\, A = \left[ H_S ,A\,\right]\\
&{{\mathcal H}'}_{SE}(t)\,A = \left[ e^{i H_S t}H_{SE}(t)e^{-i H_S t} , A\,\right] .
\end{align}
Note that the sign in the exponential $e^{i t \mathcal H_S}$ is $+$ rather than $-$, because this is the evolution superoperator in the Heisenberg picture. The average of the time-ordered exponential can be expressed in terms of a cumulant generating functional for stochastic superoperator $\mathcal H'_{SE}$:
\begin{equation}\label{cumulant_def}
\overline{\mathcal T \exp\left( i\int_0^t dt' {\mathcal{H}'}_{SE}(t')\right)}\equiv\mathcal T \exp \mathcal \sum_{m}\mathcal{K}_{m} .
\end{equation}
Here $\mathcal{K}_{m}$ is the so-called cumulant \cite{Kubo_JPSJ62, VanKampen_P74_1, VanKampen_P74_2} of order $m$ in powers of $\mathcal H'_{SE}$. The cumulants can be expressed in terms of moments by expanding left and right-hand sides of Eq.~(\ref{cumulant_def}) and equating the terms of equal order. By doing so we can identify that the first cumulant equals $i\int_0^t dt' \overline{{\mathcal{H}'}_{SE}(t')}$, which vanishes because all $\overline{\omega^{(n)}_i(t)}=0$. The second cumulant is given in terms of the correlation functions of the noise:
\begin{equation}\label{eq:2nd_cumulant}
\mathcal{K}_2 = -\sum_{n,m}\sum_{i,j}\int_0^t\!\!\!\!dt_1\!\!\int_0^{t_1}\!\!\!\!\!dt_2\, \overline{\widetilde\omega^{(n)}_i(t_1)\widetilde\omega^{(m)}_j(t_2)}\mathcal J^{(n)}_i\mathcal J^{(m)}_j .
\end{equation}
Here $\widetilde{\boldsymbol\omega}^{(n)}(t)=\hat R(\Omega t|z) \cdot \boldsymbol\omega^{(n)}(t)$ is a fluctuating field transformed by means of the rotation matrix $\hat R$ into the reference frame rotating around the constant field $\Omega$ pointing in $z$ direction \cite{Szankowski_PRE13}. The superoperators $\mathcal J_i^{(n)}$ are given by
\begin{equation}
\mathcal J_i^{(n)}A = [ J_i^{(n)}, A ] .
\end{equation}
In the general case, the superoperators $\mathcal H_{SE}'(t)$ at different times do not commute, hence the higher order cumulants are given by extremely complicated expressions \cite{Kubo_JPSJ62, VanKampen_P74_1,VanKampen_P74_2, Fox_JMP74, Fox_JMP76, Aihara_PRA90}. On the other hand, if it happens that the time-ordering is not needed, all cumulants of $\mathcal H_{SE}'$ would be determined by the cumulants of $\boldsymbol\omega^{(n)}$. But these processes are Gaussian and their defining feature is that all cumulants beyond the second simply vanish. For these reasons we adopt the approximation
\begin{equation}\label{evo_superop_approx}
\mathcal U \approx e^{i t\mathcal H_S}\mathcal T \exp\mathcal{K}_2 .
\end{equation}
In Secs.~\ref{sec:transverse} and \ref{sec:phase} we consider various  noise configuration cases. For each scenario we will discuss the applicability of this approximation and calculate the explicit formula for $\mathcal U$. 

Note that while the polarization vector representation, in which ${\varrho}$ is written as a linear combination of unity and $15$ generators of $SU(4)$ group, can be used to analyze possible types of entanglement evolution \cite{Zhou_IJMPB12}, the spherical tensor operator representation used here is particularly suited for analyzing the way in which the spatial symmetries of both the two-qubit state and the noises affect entanglement dynamics. The spherical tensors form a natural basis for this problem because the outcomes of the action of the superoperators $\mathcal J^{(n)}_i$ on the initial operators are easy to derive using relations (\ref{def_tensor_m}) and (\ref{def_tensor_l}).

\section{Phase noise}  \label{sec:phase}

First we investigate the effects of pure dephasing -- the case when the noise is applied only along the quantization axis of qubits. The dephasing Hamiltonian is given by
\begin{equation}
H_{\text{deph}}=\Omega(J^{(1)}_z+J^{(2)}_z)+\omega^{(1)}_z(t)J^{(1)}_z+\omega^{(2)}_z(t)J^{(2)}_z .
\end{equation}
To understand the relation between the symmetry of the state and the degree of noise correlation, we examine in detail how the decoherence affects entanglement.

According to Eq.~(\ref{bell_diag_conc}) the concurrence of any $X_{\mathrm{corr}}$-state is given in terms of the expectation values of the spherical tensor operators, which can be propagated in the Heisenberg picture. In the case of dephasing noise, the Hamiltonian commutes at different times and the time-ordering in (\ref{evo_superop}) is not required. Hence, Eq.~(\ref{evo_superop_approx}) is exact and the evolution superoperator for pure phase noise becomes
\begin{align}
\mathcal U_{\mathrm{deph}} {}& = e^{ i\Omega t\left( \mathcal J_z^{(1)}+\mathcal J_z^{(2)}\right)} \, e^{-  \Gamma_{1}(t)(\mathcal J_z^{(1)})^2 - \Gamma_{2}(t)(\mathcal J_z^{(2)})^2  }
\nonumber\\
& \times e^{-\Gamma_{\times}(t)\left(\mathcal J_z^{(1)}\mathcal J_z^{(2)}+\mathcal J_z^{(2)}\mathcal J_z^{(1)}\right)} ,
\end{align}
where the decay functions $\Gamma_{n}(t)$ and $\Gamma_{\times}(t)$ are given by
\begin{align}
\Gamma_{n}(t) & = \int_0^t\!\!\!dt'\int_0^{t'}\!\!\!dt'' \kappa^{(n,n)}_{zz}(t'-t'') , \label{eq:Gamma_n}\\
\Gamma_{\times}(t) & = \int_0^t\!\!\!dt'\int_0^{t'}\!\!\!dt'' \kappa^{\times}_{zz}(t'-t'')  .\label{eq:Gamma_x}
\end{align}
When the noise has a well-defined autocorrelation time, $t_{c}$, then for $t\! \ll \! t_{c}$, $\Gamma_{n}(t) \! \approx \frac{1}{2} \sigma^{2}_{n}t^2$ and $\Gamma_{\times}(t) \! \approx \frac{1}{2} \sigma^{2}_{\times}t^2$, where $\sigma^{2}_{n} \! \equiv \! \kappa^{(n,n)}_{zz}(0)$ and $\sigma^{2}_{\times} \! \equiv \! \kappa^{\times}_{zz}(0)$ are the total powers of the respective noises. When $\sigma^{2}_{n}$ and $\sigma^{2}_{\times}$ are large enough, most of the coherence decay occurs in this regime, and the decay is Gaussian, $\sim \! \exp(-(t/T_{2}^{*})^2)$, with decay timescale $T_{2}^{*} = \sqrt{2}/\sigma$. This is the limit in which the quasi-static bath approximation (QSBA)  holds \cite{Taylor_QIP06}: the entanglement decays not due to the presence of fluctuations during each instance of two-qubit system evolution, but due to averaging over the stochastic contributions to qubit splittings that are \emph{static} during each evolution. In other words, the dynamics of the environment occurs \emph{between} repetitions of the cycle of qubit initialization, evolution, and measurement. On the other hand, in the opposite limit of $t\! \gg \! t_{c}$, using the fact that $\kappa(\tau)$ is non-negligible only for $|\tau| \leq t_{c}$ , 
double integrals in Eqs.~(\ref{eq:Gamma_n}) and (\ref{eq:Gamma_x}) can be approximated by 
$\Gamma_{n/\times}(t) \! \approx \! t/T^{n/\times}_{2}$, i.e., we recover the Markovian decoherence limit.  Hence, we obtain the Bloch-Redfield theory for the case of dephasing \cite{Blum}. The characteristic decay timescales are then given by
\begin{subequations} \label{eq:rates}
\begin{equation} \label{eq:T2}
\frac{1}{T^{n}_{2}} \equiv \frac{1}{2}\int_{-\infty}^{\infty} \kappa^{(n,n)}(\tau)\text{d}\tau = S_{n}(\omega = 0) ,
\end{equation}
\begin{equation} \label{eq:T2'}
\frac{1}{T^{\times}_{2}} \equiv \frac{1}{2}\int_{-\infty}^{\infty} \kappa^{\times}(\tau)\text{d}\tau = S_{\times}(\omega = 0) ,
\end{equation}
\end{subequations}
where $S_{n/\times}(\omega)\!= \! \int\kappa^{(n,n)/\times}(\tau)e^{i\omega \tau}\text{d}\tau$ is the spectral density of either the noise felt by the $n$-th qubit, or the spectrum of cross-correlation of noises felt by the two qubits. As an example, we take the noises to be Ornstein-Uhlenbeck (OU) processes \cite{Wang_RMP45}, for which $\kappa(\tau) = \sigma^{2}e^{-|\tau|/t_c}$. Hence, $\Gamma(t) = \sigma^{2}t_{c}[ t - t_{c}(1-e^{-t/t_{c}})]$, from which both of the above limits can be derived without difficulty.

The action of $\mathcal U_{\mathrm{deph}}$ can be easily determined because, according to Eq.~(\ref{def_tensor_m}), spherical tensor operators are eigen-operators of $\mathcal J_z^{(n)}$. Hence, the evolution of operators in the Heisenberg picture are such that:
\begin{align}
& T^{(1)}_{l_1 m_1}\otimes T^{(2)}_{l_2 m_2} \xrightarrow{\mathcal U_{\text{deph}}} e^{i(m_1+m_2)\Omega t}\times\nonumber\\
&\times e^{-\left[ m^{2}_{1}\Gamma_{1}(t) + m^{2}_{2}\Gamma_{2}(t) + 2m_{1}m_{2}\Gamma_{\times}(t) \right]}  \,T^{(1)}_{l_1 m_1}\otimes T^{(2)}_{l_2 m_2} \,\, .\label{H_pic_deph}
\end{align}
Since the spherical tensors products $T^{(1)}_{1 m_1}\otimes T^{(2)}_{1 m_2}$ are eigen-operators of the evolution, $X_{\mathrm{corr}}$--states remain in their class and Eq.~(\ref{bell_diag_conc}) for the concurrence is valid at all times.  Thus, within the Heisenberg picture, using Eq.~(\ref{H_pic_deph}) and the previously calculated expectation values (\ref{eq:S_exp_vals}), (\ref{eq:T_exp_vals}) and (\ref{eq:NOON_exp_vals}), it is now easy to determine the concurrence of the evolving Bell states. By noting that $T^{(1)}_{10}\otimes T^{(2)}_{10}$ remains constant, we see that the general formula (\ref{bell_diag_conc}) simplifies to,
\begin{align}
\mathcal C_{\Psi_\pm}(t) &= 4 \big|\big\langle T^{(1)}_{11}\otimes T^{(2)}_{1\,-1}\big\rangle_{\varrho_{\Psi_\pm}(t)}\big|=e^{-2[\Gamma(t)-\Gamma_{\times}(t)]}  ,\label{eq:dephas_S_and_T} \\
\mathcal C_{\Phi_\pm}(t) &=4 \big|\big\langle T^{(1)}_{11}\otimes T^{(2)}_{11}\big\rangle_{\varrho_{\Phi_\pm}(t)}\big|= e^{-2[\Gamma(t)+\Gamma_{\times}(t)]} \,\, , \label{eq:dephas_NOON}
\end{align}
where we have used the notation $\mathcal C_X$ for the concurrence of the state $\varrho_X(t)$ ($X=\Psi_\pm,\Phi_\pm$), and we have defined $\Gamma(t) \equiv [\Gamma_{1}(t)+\Gamma_{2}(t)]/2$.

This simple case of decoherence  already reveals the strong interplay between the degree of correlation of the noise and the symmetry of the states. For completely decorrelated noises, i.e., $\Gamma_{\times} = 0$, all four Bell states lose their initial entanglement at exactly the same rate, $e^{-2\Gamma(t)}$. In this case the evolution of spherical tensor operators reads
\begin{align}
 T^{(1)}_{l_1 m_1}\otimes T^{(2)}_{l_2 m_2}{} &\xrightarrow[\Gamma_\times=0]{\mathcal U_{\text{deph}}} e^{i(m_1+m_2)\Omega t}\times\nonumber\\
&\times e^{-m_1^2 \Gamma_1(t)}e^{-m_2^2 \Gamma_2(t)}\, T^{(1)}_{l_1 m_1}\otimes T^{(2)}_{l_2 m_2} .
\end{align}
It is evident that noises decohere each qubit independently -- the constituent tensor operators decay at the rate set by their respective magnetic numbers $m_n$ and the {\it total magnetic number} $M=m_1+m_2$ is of no importance. As a result, expectation values of $T^{(1)}_{11} \otimes T^{(2)}_{11}$ and $T^{(1)}_{11} \otimes T^{(2)}_{1-1}$ decay at identical rates.

As the degree of correlations between noises increases, the decay for $\ket{\Phi_\pm}$ states accelerates while that for $\ket{\Psi_\pm}$ slows and comes to a complete halt in the limit of fully correlated noise, in which $\Gamma(t) = \Gamma_{\times}(t) = \Gamma_{1}(t) = \Gamma_{2}(t)$. The evolution of spherical tensor operators for the case of fully correlated noise reads:
\begin{align}
 T^{(1)}_{l_1 m_1}\otimes T^{(2)}_{l_2 m_2} {}&\xrightarrow[\Gamma_\times=\Gamma_{1,2}]{\mathcal U_{\text{deph}}} e^{i(m_1+m_2)\Omega t}\times\nonumber\\
&\times e^{-(m_1+m_2)^2 \Gamma(t)} \,T^{(1)}_{l_1 m_1}\otimes T^{(2)}_{l_2 m_2}.
\end{align}
Note that in this case both the deterministic field and the phase noise couple to the total magnetic number $M=m_1+m_2$. The entanglement of the $\ket{\Psi_\pm}$ states is not affected by fully correlated phase noise since their concurrence is set by unpolarized tensor product $T^{(1)}_{11}\otimes T^{(2)}_{1\,-1}$ with total magnetic number $M=0$. Thus, we recover the well known result that $\ket{\Psi_\pm}$ states span a decoherence-free subspace \cite{Duan_PRA98, Lidar_ACP14} with respect to fully correlated dephasing noise. On the other hand, the $\ket{\Phi_\pm}$ states are most sensitive to such a noise. The concurrence of these states is given by maximally polarized tensors $T^{(1)}_{11}\otimes T^{(2)}_{11}$ with $M=2$, which are damped by the noise most severely.

In summary, the phase noises couple to the magnetic quantum numbers  $m_1$ and $m_2$ as well as to total $M=m_1+m_2$. Depending on the degree of the correlations between noises acting on qubits the role of $M$ is either diminished or enhanced with respect to the role of each individual $m_1$ and $m_2$.

\section{Beyond pure dephasing: the influence of transverse noise}  \label{sec:transverse}

\subsection{Isotropic white noise}
The addition of the noise in the plane perpendicular to the $z$ axis makes the decoherence sensitive not only to the magnetic numbers, but also to the length of the angular momenta of the constituents as well as the total angular momentum. In order to investigate the role of the full rotational symmetry of the states, we focus now on isotropic noise,
\begin{equation}
H_{\text{iso}} = \Omega(J^{(1)}_z+J^{(2)}_z) + \boldsymbol{\omega}^{(1)}(t)\cdot{\mathbf J}^{(1)}+\boldsymbol{\omega}^{(2)}(t)\cdot{\mathbf J}^{(2)} ,
\end{equation}
where the orthogonal components of the noise vectors are independent (i.e., $\kappa_{ij} \! \propto \! \delta_{ij}$). With $\kappa_{ii}(t) \!= \! \kappa(t)$ for $i= x$, $y$, and $z$, the statistical properties of Gaussian vector processes $\boldsymbol\omega^{(n)}(t)$, $n = 1,2$, are rotationally invariant, and the noises are isotropic. Finally, in order to keep the discussion as transparent as possible, we assume that the noise is white, $\kappa(t-t') = (2/T) \, \delta(t-t')$. In this case the degree of cross-correlation of noises is given by a single parameter $\gamma\in [0,1]$ [see Eqs.~(\ref{eq:knn_white}) and (\ref{eq:kx_white})]. The assumption of vanishing (in comparison with the characteristic timescales of the dynamics of the qubits) correlation time makes the second-order cumulant expansion from Eq.~(\ref{evo_superop_approx}) exact \cite{Fox_JMP74}. The discussion of the situation in which the noise has a finite correlation time is much more involved (see Sec.~\ref{sec:transverse_nonmarkovian}).

With the above assumptions the explicit expression for evolution superoperator is given by
\begin{equation}
\mathcal U_{\mathrm{iso}} = e^{i\Omega t\mathcal J_z}e^{-\frac {t}{T}\left[ \gamma\boldsymbol{\mathcal J}^2 + (1-\gamma)\left(\boldsymbol{\mathcal J}_1^{2}+\boldsymbol{\mathcal J}_2^{2}\right)\right]} .
\end{equation}
Here we introduce the total spin superoperators $\mathcal J_i = \mathcal J_i^{(1)}+\mathcal J_i^{(2)}$ and superoperators of square of length of angular momentum $\boldsymbol{\mathcal J}^2 =\sum_i \mathcal J_i^2$ as well as $\boldsymbol{\mathcal J}_n^{2}=\sum_i\mathcal (\mathcal J_i^{(n)})^2$. The action of these superoperators is derived straightforwardly form equations (\ref{def_tensor_m}) and (\ref{def_tensor_l}):
\begin{align}
&\boldsymbol{\mathcal J}_n^{2}\,T^{(n)}_{l m} = l(l+1)T^{(n)}_{l m} , \\
&\boldsymbol{\mathcal J}^2 T_{LM(l_1 l_2)} = L(L+1)T_{LM(l_1 l_2)} , \\
&\boldsymbol{\mathcal J}_n^{2}\, T_{LM(l_1 l_2)} = l_n(l_n+1)T_{LM(l_1 l_2)} ,
\end{align}
where $T_{LM(l_1 l_2)}$ is a total angular momentum tensor defined by Eq. (\ref{ang_mom_addition}).

Using these results, we can see how the coupling to angular momenta affects the evolution. First let us consider the  fully correlated noise limit in which
\begin{equation}
T_{L M(l_1 l_2)} \xrightarrow[\gamma=1]{\mathcal U_{\mathrm{iso}}} e^{i M \Omega t} e^{-L(L+1)\frac tT}T_{L M(l_1 l_2)}.
\end{equation}
The fully correlated noise couple to the {\it total angular momentum} $L$, while the magnetic number $M$ and the angular momenta of the constituents $l_1$, $l_2$ are irrelevant. On the other hand, for independent noises, the evolution yields
\begin{equation}
T_{LM(l_1 l_2)} \xrightarrow[\gamma=0]{\mathcal U_{\text{iso}}} e^{i M \Omega t}
 e^{-l_1(l_1+1)\frac tT}e^{-l_2(l_2+1)\frac tT}T_{LM(l_1l_2)} ,
\end{equation}
and also
\begin{align}
T^{(1)}_{l_1 M-m}{}&\!\otimes T^{(2)}_{l_2 m} \xrightarrow[\gamma=0]{\mathcal U_{\text{iso}}} e^{i M \Omega t} \times\nonumber\\
&\times e^{-l_1(l_1+1)\frac tT}e^{-l_2(l_2+1)\frac tT}T^{(1)}_{l_1 M-m}\!\otimes T^{(2)}_{l_2 m} .\label{iso_g0_prod}
 \end{align}
Since the noises couple independently to the angular momenta of the constituents $l_1$, $l_2$ instead of to $L$, the global symmetry of the operator does not affect the evolution. Consequently, both total angular momentum tensors as well as tensor products are eigen-operators of $\mathcal U_{\mathrm{iso}}$.

Finally, in the general case of partially correlated noises, we find
\begin{align}
&T_{L M(l_1 l_2)}\xrightarrow[0<\gamma<1]{\mathcal U_{\mathrm{iso}}} e^{i M \Omega t}\times\nonumber\\
&e^{-\gamma L(L+1)\frac tT} e^{-(1-\gamma)[l_1(L_1+1)+l_2(l_2+1)]\frac tT}T_{L M(l_1 l_2)}.\label{total_tensor_H_pic}
\end{align}
The fact that total angular momentum spherical tensors are eigen-operators of the evolution for all values of $\gamma$ guarantees that Bell states remain in the class of $X_{\mathrm{corr}}$--states and Eq.~(\ref{bell_diag_conc}) holds at all times. The expectation values of spherical tensor products which contribute to concurrence can be calculated by inverting relation (\ref{ang_mom_addition}), thus expressing products in terms of linear combinations of tensors $T_{LM(l_1 l_2)}$ and evolving them according to Eq. (\ref{total_tensor_H_pic}). Following these steps, we obtain, utilizing the decompositions of Bell states, the following explicit formulae for concurrence:
\begin{align}
&\mathcal C_{\Psi_-}(t) = \text{max}\left\{ 0 , -\frac 12 + \frac 32 e^{-4(1-\gamma)\frac tT} \right\}\label{C_S_iso},\\
&\mathcal C_{\Psi_+} (t) = \mathcal C_{\Phi_\pm}(t)=\nonumber\\
 &  =\text{max}\Big\{ 0 , -\frac 12 + \frac 16 e^{-4(1-\gamma)\frac tT}\left(1 +  8\,e^{-6\gamma \frac tT}\right)\Big\} . \label{C_T_iso}
\end{align}

\begin{figure}
\centering
\includegraphics[scale=.9]{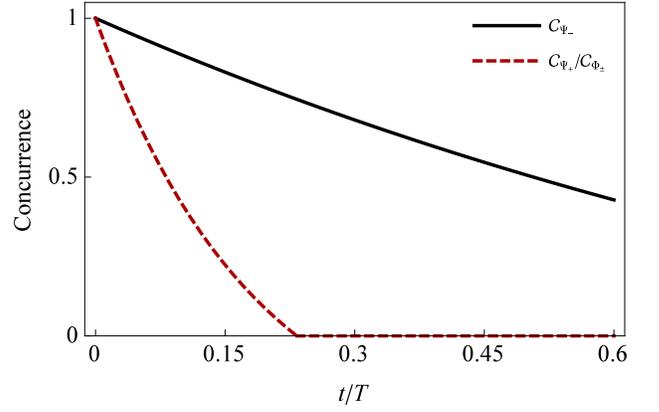}
\caption{The evolution of concurrences of Bell states affected by partially correlated isotropic white noises. The degree of noise cross-correlation is $\gamma=0.8$. The concurrence of $|\Psi_-\rangle$ state is given by Eq.~(\ref{C_S_iso}) (solid black) and the concurrence of states $|\Psi_+\rangle$ and $|\Phi_\pm\rangle$ is given by Eq.~(\ref{C_T_iso}) (red dashed).}\label{fig2}
\end{figure}

We note that, as in the case of phase noise, independent noises destroys entanglement for all four Bell states at the same rate:
\begin{equation}\label{eq:conc_iso_independent}
\gamma=0:\;\mathcal C_{\Psi_\pm}(t) =\mathcal C_{\Phi_\pm}(t) = \text{max}\left\{ 0,
-\frac 12 + \frac 32 e^{-4\frac tT}
\right\}  .
\end{equation}
As the degree of correlation between the noises increases, the global symmetries of the states become more and more important, and the entanglement decay rates of Bell states start to differ. In particular, for strongly correlated noises, the $\ket{\Psi_-}$ state retains its entanglement for times much longer than the $\ket{\Psi_+}$ and the $\ket{\Phi_\pm}$ states (see Fig.~\ref{fig2}). In the limiting case of fully correlated noise, the concurrence reads
\begin{equation}\label{g_1_iso_conc}
\gamma=1: \left\{ \begin{array}{l}
\mathcal C_{\Psi_-}(t) =1, \\[.3cm]
\mathcal C_{\Psi_+}(t) =\mathcal C_{\Phi_\pm} (t) = \mathrm{max}\Big\{ 0 ,\,-\frac 13 +\frac 43 e^{-6\frac tT} \Big\} 
\end{array} \right. .
\end{equation}
The immunity of the $\ket{\Psi_-}$ state to correlated noise can be explained as follows. As mentioned above, for $\gamma \sim 1$, the noise mostly couples to the total angular momenta. The $\ket{\Psi_-}$ state is a scalar, i.e., it is spanned only by the unit operator and the $L=0$ spherical tensor. Hence, the fully correlated noise has nothing to couple to and strongly correlated noise has only a weak effect. On the other hand, $\ket{\Psi_+}$ and $\ket{\Phi_\pm}$ states have a total spin of one, and the tensor operators with $L>0$ are present in their decompositions. The noise can couple to these higher $L$ and the states are not protected. Moreover, since isotropic noise ignores the polarization of the states quantified by the magnetic quantum numbers  $M$ of the spherical tensors appearing in their expansion, the entanglement of the states $\ket{\Psi_+}$ and $\ket{\Phi_\pm}$ suffer decoherence in the same way.

\subsection{Purely transverse white noise}

\begin{figure}
\centering
\includegraphics[scale=.9]{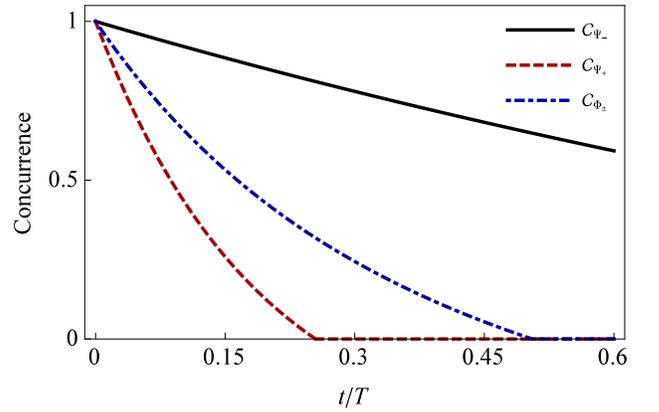}
\caption{The evolution of concurrences of Bell states affected by partially correlated transverse white noises. The degree of noise cross-correlation is $\gamma=0.8$. The concurrence of $|\Psi_-\rangle$ state is given by Eq.~(\ref{eq:C_S_tr}) (solid black), the concurrence of $|\Phi_\pm\rangle$ is given by Eq.~(\ref{eq:C_NOON_tr}) (blue dot-dashed) and the concurrence of $|\Psi_+\rangle$ is given by Eq.~(\ref{eq:C_T_tr}) (red dashed).}\label{fig3}
\end{figure}

In the previous section we saw that isotropic noise couples only to angular momentum, ignoring the magnetic numbers of spherical tensors. It is instructive to examine the case where full spherical symmetry is lifted, and only fluctuations in the plane perpendicular to the quantization axis remains. We assume the following form of the Hamiltonian:
\begin{equation}
H_{\mathrm{tr}} =  \Omega(J^{(1)}_z\!+\!J^{(2)}_z)+\!\!\! \sum_{n=1,2}\!\!( \omega_x^{(n)}(t) J_x^{(n)}\!+ \omega_y^{(n)}(t) J_y^{(n)}) .
\end{equation}
The noise correlation functions are taken to be of the same form as in the previous case, $\kappa_{ij} \! \propto \! \delta_{ij}$, with $\kappa_{ii} = \kappa$, but now only with $i=x,y$. Statistical independence of the orthogonal components of the noise results in cylindrical symmetry of the problem. As in the previous case of isotropic noises this time we also adopt Markovian approximation $\kappa(t-t')=\frac 2T\delta(t-t')$ in order to keep the physical picture as transparent as possible. With the above assumptions the evolution superoperator is exactly given by (\ref{evo_superop_approx}) and it reads
\begin{align}
&\mathcal U_{\mathrm{tr}}=e^{i\Omega t\mathcal J_z}e^{-[(1-\gamma)(\boldsymbol{\mathcal J}_1^{2}+\boldsymbol{\mathcal J}_2^{2})-\mathcal J_z^2]\frac {t}T}\times\nonumber\\
&\times\exp\left\{-\left[ \gamma \boldsymbol{\mathcal J}^2 +(1-\gamma)(\mathcal J_z^{(1)}\mathcal J_z^{(2)}+\mathcal J_z^{(2)}\mathcal J_z^{(1)})\right]\frac{t}T\right\}. \label{tr_evo_superop}
\end{align}
Because $[\boldsymbol{\mathcal J}^2 ,\mathcal J_z^{(1)}\mathcal J_z^{(2)}+\mathcal J_z^{(2)}\mathcal J_z^{(1)}] \neq 0$, the evolution superoperator is diagonal neither in the basis of total angular momentum spherical tensors nor in the basis of spherical tensor products. In Appendix \ref{app2} we show that the eigen-operators of $\mathcal U_{\mathrm{tr}}$ belong to the subset spanning the class of $X_{\mathrm{corr}}$--states. Therefore, evolution of the Bell states does not remove the states from the class of $X_{\mathrm{corr}}$--states, hence Eq.~(\ref{bell_diag_conc}) for the concurrence remains valid; we find
\begin{align}
&\mathcal C_{\Psi_-}(t) = \mathrm{max}\Big\{ 0 \,, -\frac 12 +\nonumber\\
&e^{-3(1+\eta)\frac tT}\frac{9\eta-(1+8\gamma)}{12\eta}+e^{-3(1-\eta)\frac tT}\frac{9\eta+(1+8\gamma)}{12\eta}\Big\} , \label{eq:C_S_tr}\\
&\mathcal C_{\Psi_+}(t) = \mathrm{max}\Big\{ 0 \,, -\frac 12 +\nonumber\\
&e^{-3(1+\eta)\frac tT}\frac{9\eta-(1-8\gamma)}{12\eta}+e^{-3(1-\eta)\frac tT}\frac{9\eta+(1-8\gamma)}{12\eta} \Big\} , \label{eq:C_T_tr}\\
&\mathcal C_{\Phi_\pm}(t) = \mathrm{max}\Big\{ 0\, ,-\frac 12  +\nonumber\\
&e^{-2\frac tT}+e^{-3(1+\eta)\frac tT}\frac{3\eta-1}{12\eta}+e^{-3(1-\eta)\frac tT}\frac{3\eta+1}{12\eta}\Big\} . \label{eq:C_NOON_tr}
\end{align}
Here $\eta = \sqrt{1+8 \gamma^2}/3$, which ranges from $\eta =1/3$ for $\gamma=0$ to $\eta=1$ for $\gamma=1$. The concurrences are drawn as a function of time in Fig.~\ref{fig3}.

The most evident qualitative difference between isotropic and transverse noise is the disparity in the disentanglement rate of $\ket{\Psi_+}$ and $\ket{\Phi_\pm}$ states [see Eq.~(\ref{C_T_iso})]. The source of this difference can be traced back to the dependence of decoherence on the polarizations of states which results from the symmetry of the transverse noises.

As in previous cases, in the limit of independent noise all four Bell states decohere at the same rate,
\begin{equation}
\gamma=0:\mathcal C_{\Psi_\pm}(t) =\mathcal C_{\Phi_\pm} (t) = \mathrm{max}\Big\{ 0\, , -\frac 12 +\frac{e^{-2\frac tT}}2(2+e^{-2\frac  tT})\Big\} .
\end{equation}
The rate of entanglement decay is slower than in the case of independent isotropic noise. To explain this difference we examine the evolution of spherical tensor products, which become eigenoperators of $\mathcal U_{\mathrm{tr}}$ for $\gamma=0$:
\begin{align}
&T^{(1)}_{l_1 m_1}\otimes T^{(2)}_{l_2 m_2}\xrightarrow[\gamma=0]{\mathcal U_{\mathrm{tr}}} e^{i(m_1+m_2)\Omega t }\times\nonumber\\
&\; \; e^{-[l_1(l_1+1)-m_1^2]\frac tT}e^{-[l_2(l_2+1)-m_2^2]\frac tT}T^{(1)}_{l_1 m_1}\otimes T^{(2)}_{l_2 m_2} . \label{cyl_g0_prod}
\end{align}
In a fashion similar to the isotropic case, transverse noise couples to angular momenta $l_1$ and $l_2$ of the constituting spherical tensors (see Eq. (\ref{iso_g0_prod})). The difference between the two lies in the dependence of decay rates on magnetic quantum numbers  $m_1$ and $m_2$. In general, contribution from magnetic quantum numbers  blunt the effects of transverse noise, while they are completely irrelevant in the case of isotropic noise.

Next, for fully correlated noise ($\gamma=1$) the scalar state $\ket{\Psi_-}$ is unaffected by decoherence, while  $\ket{\Psi_+}$ and $\ket{\Phi_\pm}$ states disentangle
\begin{equation}
\gamma=1: \left\{ \begin{array}{l}
\mathcal C_{\Psi_-}(t) = 1 , \\[.2cm]
\mathcal C_{\Psi_+}(t) = \mathrm{max}\Big\{ 0 \,, -\frac 13 +\frac 43 e^{-6\frac tT}\Big\} , \\[.2cm]
\mathcal C_{\Phi_\pm}(t) = \mathrm{max}\Big\{ 0\, ,-\frac 13  +\frac {e^{-2\frac tT}}3\left(3+e^{-3\frac tT}\right)\Big\} .
\end{array} \right.
\end{equation}
Note that the rate of disentanglement of $\ket{\Psi_+}$ state is the same as in the case of fully correlated isotropic noises [see Eq.~(\ref{g_1_iso_conc})]. For $\gamma=1$ the evolution superoperator $\mathcal U_{\mathrm{tr}}$ simplifies and becomes diagonal in the basis of total angular momentum tensors $T_{LM(l_1 l_2)}$:
\begin{equation}  \label{Eq:TLM}
T_{LM(l_1 l_2)}\xrightarrow[\gamma=1]{\mathcal U_{\mathrm{tr}}} e^{iM\Omega t }e^{-[ L(L+1)-M^2 ]\frac tT}T_{LM(l_1 l_2)} .
\end{equation}
As in all the previous scenarios $\ket{\Psi_-}$, which is spanned by the $T_{00(11)}$ tensor, is unaffected by fully correlated noise. The state $\ket{\Psi_+}$ is spanned by scalar tensor $T_{00(11)}$ as well, but also by the unpolarized tensor $T_{20(11)}$ with $M=0$. This means that the evolution of this state is unaffected by the presence of noise in $z$ direction and therefore there is no difference between evolution induced by fully correlated transverse and isotropic noise. On the other hand, the $\ket{\Phi_\pm}$ states contain maximally polarized parts $T_{2\pm 2(11)}$ which decay slower than the unpolarized tensor because of the $-M^2$ term on the right hand side of Eq.~(\ref{Eq:TLM}), which is absent for isotropic noise.

The noise in transverse directions also brings in additional important effect. In the presence of pure phase noise the entanglement of our states of interest vanished only asymptotically [see Eqs.~(\ref{eq:dephas_S_and_T}) and (\ref{eq:dephas_NOON})]. With transverse noise present, the second argument of $\text{max}$ function in formulas for concurrences, Eqs.~(\ref{eq:C_S_tr}) and (\ref{eq:C_NOON_tr}) [as well as (\ref{C_S_iso}) and (\ref{C_T_iso}) in case of isotropic noise] has the form of $-1/2$ plus exponentially decaying positive terms. Clearly, at some point in time, these expressions becomes negative. Hence, for all four Bell states, there is such a time $t^{(SD)}<\infty$ when the state becomes separable (with the only exception of $|\Psi_-\rangle$ subjected to fully correlated noise). The phenomenon of abrupt vanishing of the entanglement is known as {\it sudden death of entanglement} \cite{Zyczkowski_PRA01, Yu_PRL04, Yu_Science09}.

\subsection{Transverse noise with finite correlation time} \label{sec:transverse_nonmarkovian}

When the noises affecting the qubits $n = 1,2$ have finite correlation times $t^{(n)}_c$, the quantification of the influence of transverse fluctuations becomes much more complicated \cite{VanKampen_P74_1, VanKampen_P74_2,Fox_JMP74, Fox_JMP76, Budimir_JSP87, Aihara_PRA90, Makhlin_PRL04} compared to the previously discussed case of white noise. When the timescale of interest, $t$, is shorter than the noise autocorrelation time, $t \ll t^{(n)}_c$, the fact that superoperators ${\mathcal{H}'}_{SE}(t')$ in Eq.~(\ref{cumulant_def}) do not commute at different times cannot be ignored, and higher-order cumulants (given by rather complicated expressions) should be taken into account \cite{Fox_JMP74,Fox_JMP76,Aihara_PRA90}. The exact solution is impossible to obtain, and below we will only give a rather brief discussion of possible approximate analytical solutions with some results of exact simulations added to illustrate the main points. Since the topic of influence of transverse noise on coherence and entanglement dynamics is quite involved \cite{Makhlin_PRL04,Bergli_PRB06, Cywinski_PRA14,Brox_JPA12,Benedetti_IJQI14} the aim of this section is to give only a qualitative picture.

The first thing to note is the significant role that the qubit energy splitting $\Omega$ plays in this case. When qubits are interpreted as spin $1/2$ particles, then the Zeeman splitting causes precession of spin around the $z$-axis with frequency $\Omega$. If the period of the precession is large in comparison to the correlation time, $\Omega t_c\gg 1$, the fast revolution effectively averages-out the slow fluctuations and suppresses the decoherence induced by the transverse noise, while the phase noise remains unaffected. A similar principle applies to a top which does not topple if it spins fast enough. The qualitative effect of finite $\Omega$ is captured in the second-cumulant approximation where the decay functions are given by Eq.~(\ref{eq:2nd_cumulant}). The effective averaging of fluctuations described above results from the transformation of the noise vector components to the rotating frame. For example, in the case of purely transverse and fully correlated noise, the approximate evolution superoperator reads
\begin{equation}
\mathcal U_{\mathrm{tr}}^{(\text{finite $t_c$})}\approx e^{i(\Omega+\Delta\Omega)t\mathcal J_z} \, e^{-\Gamma_\perp(t)[\boldsymbol{\mathcal J}^2-\mathcal J_z^2]} .
\end{equation}
Here the decay function is given by
\begin{equation}
\Gamma_\perp(t) = \int_0^t \!\!\!\!dt'\!\!\int_0^{t'}\!\!\!\!dt''\,\kappa(t'-t'')\cos\left[\Omega(t'-t'')\right] ,
\end{equation}
and we see that it is modified by the precession of spins \footnote{{The splitting renormalization is given by $\Delta\Omega t = \int_0^t \!dt'\!\int_0^{t'}\!dt''\kappa(t'-t'')\sin\left[\Omega(t'-t'')\right]$}}.
For $t \ll t_{c}$ we have $\Gamma_{\perp}(t) \propto \sigma^{2}/\Omega^{2}$ where $\sigma^{2} = \kappa(0)$ is the total noise power (note that despite the need for a more careful treatment, the same conclusion holds for the case of $1/f^{\beta}$ noise \cite{Paladino_RMP14}, the influence of which on two-qubit entanglement dynamics was recently considered in \cite{Bellomo_PRA10,Benedetti_PRA13}). Hence, for $\Omega \gg \sigma$, we expect the influence of the transverse noise to be strongly suppressed. To be precise, this conclusion holds qualitatively for the relaxation processes of the qubits (i.e., the processes in which the diagonal elements of the two-qubit density matrix spanned by the unpolarized tensors change due to the presence of the noise), while the contribution of transverse noise to dephasing of the two qubits has to be considered more carefully. The qualitative explanation of this fact is the following. For $\Omega \gg \sigma$ it is very hard for the noise to rotate the qubits about the in-plane axes. Furthermore, the qubit energy eigenstates split by $\Omega$ so the transitions require a transfer of such energy between the qubits and the noise source, and the rate of such a transfer is $\approx S(\Omega)$, where $S(\omega)$ is the spectral density of the noise (this argument is only qualitative, since it implicitly uses the Markovian approximation, which is not valid here). Using the Ornstein-Ulhenbeck noise as an example, we have $S(\Omega) \propto \sigma^{2}/\Omega^{2}t_{c} \ll 1/t_{c}$ for $\Omega t_{c} \gg 1$, so very little relaxation can occur on a timescale of $t \ll t_{c}$. On the other hand, the processes of phase randomization can occur without any energy exchange between the qubits and the bath: virtual exchanges of energy are enough to scramble the phase relationship between the states in a superposition.

\begin{figure}
\centering
\includegraphics[scale=.9]{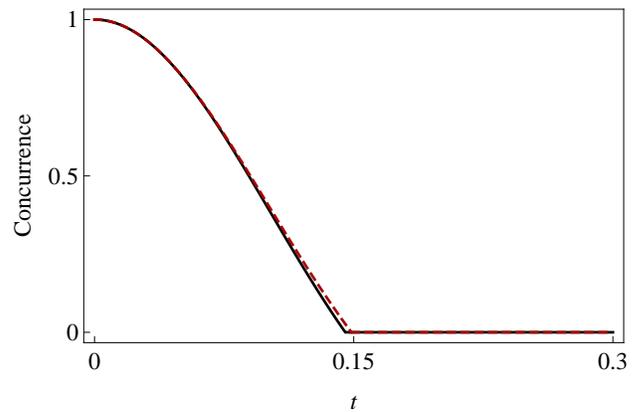}
\caption{Evolution of the concurrence of Bell states affected by independent isotropic colored noise (in this case all four Bell state disentangle at the same rate). We compare the exact numerical simulation (solid black) with the second-order cumulant approximation (red dashed). The noises are chosen to be OU processes characterized by $\kappa^{(n,n)}_{ii}(t)=\sigma_n^2 \exp\left(-\frac {|t|}{t_c^{(n)}}\right)$ with $\sigma_1=\sigma_2=5$, $t_c^{(1)}=t_c^{(2)}=10$ and the qubit splitting $\Omega=1$.}
\label{fig:QSBA_isotropic}
\end{figure}

Another thing to note is that in the presence of longitudinal noise, the influence of which is unaffected by $\Omega$, the disentanglement of the two qubits at large $\Omega$ will be simply determined by this noise, i.e., situation is as discussed in Sec.~\ref{sec:phase}. For this reason we focus now on the case of purely transverse noise. We note that in the limit of $\Omega \ll \sigma$, for strong coupling to slow noise, the second-cumulant approximation for isotropic noise describes the entanglement decay quite well. This is illustrated in Fig.~\ref{fig:QSBA_isotropic}, where the simulations of concurrence decay due to interaction with isotropic OU noise are compared with results of second order cumulant approximation.

Let us now focus on the regime $\Omega \gg \sigma$ for purely transverse noise. We neglect the effects of noise on the diagonal components of the density matrix (which amount to fluctuations of these elements having amplitudes $\sim \! (\sigma/\Omega)^2$), and focus on the dephasing caused by the noise. To second order with respect to the transverse terms in the stochastic Hamiltonian, we obtain the effective interaction:
\begin{equation}
\tilde{H}_{SE} = \sum_{n=1,2} \frac{(\omega^{(n)}_{x}(t))^2 + (\omega^{(n)}_{y}(t))^2}{2\Omega} J^{(n)}_{z} .  \label{eq:Heff}
\end{equation}

From Eq.~(\ref{eq:Heff}), the effective Hamiltonian is of pure dephasing form, so, as discussed in Sec.~\ref{sec:phase}, when using a cumulant expansion to calculate the evolution of the noise-averaged density matrix, we do not have to deal with non-commuting superoperators. However, the cumulant expansion cannot be truncated at second order: the reason is the fact that the square of the Gaussian noise, $(\omega^{(n)}_{x,y})^2$, does not have Gaussian statistics \cite{Makhlin_PRL04, Cywinski_PRA14}. In fact, in order to describe the entanglement decay beyond the short-time limit (when both the coherences and the concurrence are close to their initial values) one has to re-sum an infinite number of terms in the cumulant expansion \cite{Makhlin_PRL04,Cywinski_PRA14}. Using the results of such an infinite-order cumulant expansion, we arrive at the following evolution of the spherical tensor operators products in the case of independent noises:
\begin{align}
& T^{(1)}_{j_1 m_1}\otimes T^{(2)}_{j_2 m_2} \xrightarrow{\widetilde{H}_{SE}} e^{i(m_1+m_2)\Omega t}\times\nonumber\\
& \; \; e^{-\sum_{k=1}^{\infty} \chi^{(1)}_{k}(t)}  e^{-\sum_{k=1}^{\infty} \chi^{(2)}_{k}(t)} \,T^{(1)}_{j_1 m_1}\otimes T^{(2)}_{j_2 m_2} . \label{eq:Uindep}
\end{align}
Here
\begin{equation}
\chi^{(n)}_{k} = - m^{k}_{n}\frac{i^{k}}{k} [ R^{(n),x}_{k}(t) + R^{(n),y}_{k}(t) ] ,
\end{equation}
and
\begin{align}
R^{(n),i}_{k}(t) & = \frac{2^{k-1}}{(2\Omega)^{k}}\int_{0}^{t} \text{d}t_{1}...\int_{0}^{t} \text{d}t_{k} \kappa^{(n,n)}_{ii}(t_{12})...\kappa^{(n,n)}_{ii}(t_{k1})  , \label{eq:Rt}
\end{align}
where $i= x$, $y$ and  $t_{kl} \! \equiv \! t_{k}-t_{l}$. Note that decay functions depend on the magnetic quantum numbers  $m_n$ of the constituting operators. However, for the case of fully correlated noise, the exponent contains only a single sum over the total magnetic quantum number dependent $\chi_{k}(t)$ given by
\begin{equation}
\chi_{k} = -(m_{1}+m_{2})^{k} \frac{i^{k}}{k} [ R^{x}_{k}(t) + R^{y}_{k}(t) ] .
\end{equation}
For noise with a well-defined autocorrelation time $t_c$, the resummation of all the $\chi_{k}(t)$ for $t\! \gg \! t_{c}$ terms gives an exponential decay of the averages of the tensor operator products, and therefore of the entanglement. This is expected, since Markovian decoherence should be recovered in this limit.  A more intriguing result is obtained at short times, and for noise strong enough for the decay of coherence and entanglement to be substantial on this timescale. For $t\! \ll \! t_{c}$ we can replace $\kappa^{(n,n)}_{ii}(t_{kl})$ in Eq.~(\ref{eq:Rt}) by their values at zero time-delay, given simply by $\sigma^{2}_{n}$.  This leads to
\begin{equation}
R_{k}^{(n)} \approx \frac{1}{2}\left( \frac{\sigma^{2}_{n}t}{\Omega} \right )^{k} ,
\end{equation}
from which we find
\begin{align}
\exp[-\sum_{k=1}^{\infty} \chi^{(n)}_{k}(t) ] & = \exp \left [ \sum_{k} \frac{1}{k}(im_{n}\sigma^{2}_{n}t/\Omega)^{k} \right ] \nonumber \\
& =  \exp [ -\ln(1-im_{n}\sigma^{2}_{n}t/\Omega) ] \nonumber \\
& = \frac{1}{1-im_{n}\sigma^{2}_{n}t/\Omega} . \label{eq:Wn}
\end{align}
In the case of fully correlated noise (for which of $\sigma_{1} = \sigma_{2} = \sigma$ is fulfilled) in the same fashion we obtain
\begin{equation}
\exp[-\sum_{k=1}^{\infty} \chi_{k}(t) ] = \frac{1}{1-\frac{i(m_{1}+m_{2})\sigma^{2}t}{\Omega}} . \label{eq:Wc}
\end{equation}
These are the well-known results for the quasi-static Gaussian average over a phase which is proportional to a square of Gaussian variable \cite{Makhlin_PRL04, Cucchietti_PRA05, Falci_PRL05,Taylor_QIP06, Bellomo_PRA10, Cywinski_APPA11, Cywinski_PRA14}.
Obtaining closed-form formulas in a more general case (e.g., for $t \! \sim \! t_{c}$) is a challenging task, for which appropriate methods have to be chosen for a noise with a given spectrum. We note here that an exact solution of dephasing due to the Hamiltonian (\ref{eq:Heff}) was derived for the case of Ornstein-Uhlenbeck noise \cite{Dobrovitski_PRL09}.

From the above results we obtain that for the uncorrelated noises, the decay of concurrence for all the Bell states is given by
\begin{equation}
\mathcal C_{\Psi_{\pm},\Phi_{\pm}}(t) = \prod_{n=1,2}\frac{1}{\sqrt{1+(t/\tau_{n})^2}}  .\label{eq:Cind}
\end{equation}
where $\tau_{n} = \Omega/\sigma^{2}_{n}$.
An analogous result for concurrence was derived in \cite{Bellomo_PRA10}, only with $\sqrt{1+(t/\tau_{n})}$ replaced by $[1+(t/\tau_{n})]^{1/4}$, because there the noise only along one of the transverse axes was considered.
The same result has been recently obtained for two entangled electron spins coupled by hyperfine interaction to two separate nuclear spin baths \cite{Bragar_arXiv14}. In that case, longitudinal noise (fluctuations of the effective Overhauser field that the nuclei exert on the electron spin) has very long correlation time, and using feedback techniques one can remove its influence \cite{Shulman_NC14}. The faster (but still slow compared to qubit's decoherence time) transverse field fluctuations can still be assumed to be quasi-static and Gaussian-distributed \cite{Cywinski_APPA11,Hung_PRB13} on timescale shorter than their autocorrelation time, leading to the above QSBA result for entanglement decay.

For fully correlated noise we obtain, unsurprisingly, $\mathcal C_{\Psi_{\pm}}(t) = 1$, and
\begin{equation}
\mathcal C_{\Phi_{\pm}}(t) = \frac{1}{\sqrt{1+(t/\tau)^2}} , \label{eq:Ccom}
\end{equation}
where $\tau = \Omega/2\sigma^{2}$.
The two results can be compared when the uncorrelated noises are characterized by $\sigma_{1} = \sigma_{2} = \sigma$, for which $\mathcal C(t)=  1/[1+(t/2\tau)^{2}]$. We see that for $t < 2\sqrt{2}\tau$ the entanglement of $\ket{\Phi_{\pm}}$ states in the fully correlated noise case is smaller than in the independent noise case (as it was obtained at all times for linear coupling to phase noise in Section \ref{sec:phase}), but at longer times, when the concurrences decay as $\sim \! \tau/t$ and $\sim \! 4\tau^{2}/t^2$ in the two respective cases, the entanglement is actually larger for fully correlated noise.

\begin{figure}
\centering
\includegraphics[scale=.9]{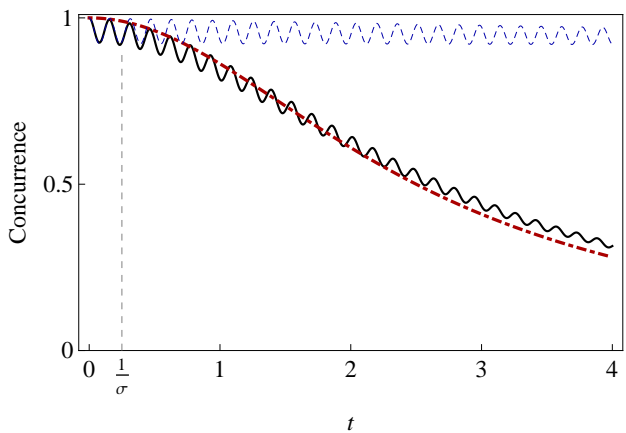}
\caption{Evolution of the concurrence of Bell states affected by independent transverse colored noises (in this case all four Bell state disentangle at the same rate). The exact numerical simulation (solid black) is compared with the second-order cumulant (blue dashed) and effective Hamiltonian (red dot-dashed) approximations. The noises are chosen to be OU processes characterized by the autocorrelation functions $\kappa^{(n,n)}_{ii}(t)=\sigma_n^2 \exp\left(-\frac {|t|}{t_c^{(n)}}\right)$ with $\sigma_1=\sigma_2=\sigma=4$, $t_c^{(1)}=t_c^{(2)}=10$ and the qubit splitting $\Omega=40$.}\label{fig:transverse_uc}
\end{figure}

\begin{figure}
\centering
\includegraphics[scale=.9]{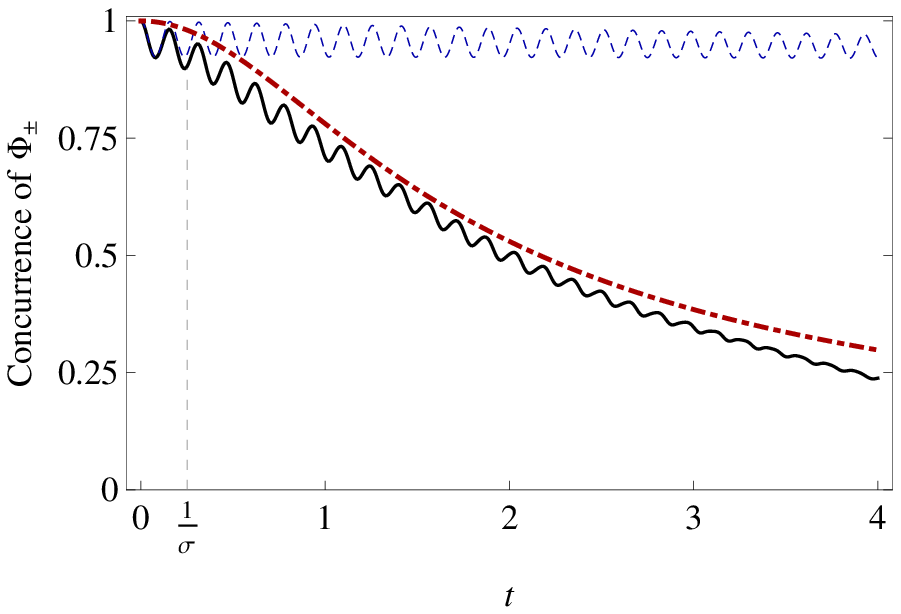}
\caption{The evolution of concurrence of the $|\Phi_{\pm}\rangle$ states under the influence of fully correlated transverse colored noises. The exact numerical simulation (solid black) is compared with the second-order cumulant (blue dashed) and effective Hamiltonian (red dot-dashed) approximations. The noises are OU processes with $\kappa^{(1,1)}_{ii}(t)=\kappa^{(2,2)}_{ii}(t)=\kappa^\times_{ii}(t)=\sigma^2\exp\left(-\frac{|t|}{t_c}\right)$ where $\sigma=4$, $t_c=10$ and $\Omega=40$.}\label{fig:transverse_cPhi}
\end{figure}

\begin{figure}
\centering
\includegraphics[scale=.9]{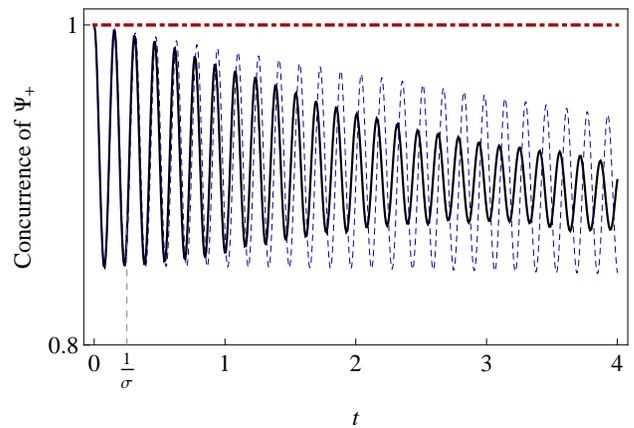}
\caption{The evolution of concurrence of the $|\Psi_+\rangle$ state under the influence of fully correlated transverse colored noises. The exact numerical simulation (solid black) is compared with the second-order cumulant (blue dashed) and effective Hamiltonian (red dot-dashed) approximations. The noises are OU processes with $\kappa^{(1,1)}_{ii}(t)=\kappa^{(2,2)}_{ii}(t)=\kappa^\times_{ii}(t)=\sigma^2\exp\left(-\frac{|t|}{t_c}\right)$ where $\sigma=4$, $t_c=10$ and $\Omega=40$.}
\label{fig:transverse_cPsi}
\end{figure}

The quality of the above pure dephasing approximation can be ascertained by looking at Figs.~\ref{fig:transverse_uc}-\ref{fig:transverse_cPsi} in which the above analytical formulas are compared with the results of exact numerical simulation (averaging the evolution of the two qubits over many realizations of OU noise) and with the results obtained with the second-order cumulant expansion method. For uncorrelated noise, Eq.~(\ref{eq:Cind}) agrees well with the simulation for $\Omega= 10 \sigma$, see Fig.~\ref{fig:transverse_uc}. The only difference between the two results is the presence of a fast small-amplitude (proportional to $(\sigma/\Omega)^2$) modulation of the exact result. On the other hand,  the second-order cumulant expansion solution exhibits the oscillatory behavior but quickly deviates from the exact result for times longer than $\sigma^{-1}$. In case of fully correlated noise and $|\Phi_\pm\rangle$ states we again see that the effective Hamiltonian approximation agrees with the exact solution with the exception of small oscillations while the cumulant approximation fails at longer times (see Fig.~\ref{fig:transverse_cPhi}). The magnitude of $\Omega$-dependent correction to the result $C_{\Psi_{\pm}}= 1$ is shown in Fig.~\ref{fig:transverse_cPsi}. With increasing $\Omega$ the amplitude of the oscillation will decay as $(\sigma/\Omega)^2$, bringing the exact $C_{\Psi_{\pm}}(t)$ closer to unity. The effective Hamiltonian approximation does not reproduce the correction since it assumes pure dephasing and the state $|\Psi_+\rangle$ is immune to fully correlated phase noise. At the same time the second-order cumulant approximation is in fairly good agreement with the numerical solution.

\section{Conclusions}  \label{sec:conclusions}

We investigated the decay of entanglement of two qubits due to exposure to classical noise when the two-qubit state is initialized to be in one of the Bell states. The cases of longitudinal noise (pure dephasing), isotropic noise, and purely transverse noise having various degrees of cross-correlation were considered. Exact results were obtained for noise with a white spectrum (autocorrelation time much shorter than characteristic decoherence times, leading to Markovian decoherence of the qubits), and for a general Gaussian noise leading to pure dephasing. The employed language of spherical tensor operators has allowed for an easy insight into relation between the rotational symmetry of a given Bell state and its subsequent decay dynamics caused by noise with a given symmetry and degree of cross-correlation. For pure dephasing, all the states disentangle in the same way for independent noises acting on the two qubits. As the degree of the cross-correlation of the noises increases, the entanglement decay of the $\ket{\Psi_{\pm}}$ states is suppressed while the decay of the entanglement of the $|\Phi_\pm\rangle$ states is being accelerated. For isotropic white noise, the $\ket{\Psi_{-}}$ state (i.e., a scalar in spherical tensor language) is unaffected by fully correlated noise.  For partially correlated noises, its entanglement decays at a slower rate than the entanglement of the three other Bell states, which decay in the same fashion. When the noise symmetry is lowered to cylindrical symmetry, in the case of purely transverse white noise we found that for finite cross-correlation of noises, the $\ket{\Psi_{-}}$ state has the smallest rate of entanglement loss, and it does not decay at all for fully correlated noise; the entanglement of the $\ket{\Phi_{\pm}}$ states decay faster, and the $\ket{\Psi_{+}}$ state is the most fragile. Again, this result can be understood by examining the decomposition of states in basis spherical tensors. Finally, we have discussed the case of transverse noise with finite autocorrelation time. The limit of strong dephasing due to slow noise (the quasi-static bath limit) was investigated for qubit energy splitting much larger than the noise amplitude using an effective Hamiltonian approach. In this case one obtains asymptotic power-law decay of entanglement and interestingly, the nonzero correlations between the noises suppresses the entanglement of $\ket{\Phi_{\pm}}$ states at short times, while enhancing it (with respect to the independent noise case) at long times.

\bigskip

\section*{Acknowledgements}
This work was supported in part by grants from the Israel Science Foundation (Grant No.~295/2011). P.~Sz. acknowledges the Foundation for Polish Science International Ph.D. Projects Program co-financed by the EU European Regional Development Fund.
{\L}C was supported by funds of Polish National Science Center (NCN), Grant No.~DEC-2012/07/B/ST3/03616.

\bigskip

\appendix
\section{Concurrence of the Bell States}\label{app1}

The density matrices of the $X_{\mathrm{corr}}$--states, written in standard two-qubit basis $\{ \left|\uparrow\downarrow\right\rangle, \left|\downarrow\uparrow\right\rangle, \left|\uparrow\uparrow\right\rangle , \left|\downarrow\downarrow\right\rangle\}$, have a form
\begin{equation}
\varrho = \left(\begin{array}{cccc}
a & 0 & 0 & w\\
0 & b & z & 0\\
0 & z^*&b &0\\
w^*&0&0&a\\
\end{array}\right) ,
\label{app:matrix}
\end{equation}
where the condition $2a+2b=1$ insures that the trace equals unity. Matrices of the form (\ref{app:matrix}) are time reversal invariant, hence the concurrence is easily evaluated because it can be expressed in term of eigenvalues of $\varrho$ itself. These eigenvalues can be determined analytically, so the concurrence is given by \cite{Yu_QIC07}
\begin{equation}\label{app:conc}
\mathcal C(\varrho)= 2\,\text{max}\left\{ 0 , |z|-a , |w|-b \right\} .
\end{equation}
Straightforward calculations show that the matrix elements $a$, $b$, $w$ and $z$ can be written in terms of the expectation values of products of the spherical tensor operators,
\begin{align}
z{}&=-2\langle T^{(1)}_{11}\otimes T^{(2)}_{1{-1}}\rangle_{\varrho} =-2\langle T^{(1)}_{1-1}\otimes T^{(2)}_{11}\rangle^*_{\varrho} , \\
w{}&=2\langle T^{(1)}_{11}\otimes T^{(2)}_{11}\rangle_{\varrho} =2\langle T^{(1)}_{1-1}\otimes T^{(2)}_{1-1}\rangle^*_{\varrho} , \\
a{}&=  \frac 14 +\langle T^{(1)}_{10}\otimes T^{(2)}_{10}\rangle_{\varrho} , \\
b{}&=  \frac 14 -\langle T^{(1)}_{10}\otimes T^{(2)}_{10}\rangle_{\varrho} .
\end{align}
Thus we obtain formula (\ref{bell_diag_conc}).\\

\section{Diagonalization of the Evolution Superoperator for the case of Transverse Noise}\label{app2}

The evolution superoperator for transverse white noise is given by Eq.~(\ref{tr_evo_superop}):
\begin{equation}
\mathcal U_{\mathrm{tr}}=e^{i\Omega t\mathcal J_z}e^{-[(1-\gamma)(\boldsymbol{\mathcal J}_1^{2}+\boldsymbol{\mathcal J}_2^{2})-\mathcal J_z^2]\frac {t}T}e^{-\frac tT\mathcal L} ,
\end{equation}
where we introduced the superoperator $\mathcal L$ defined as
\begin{equation}
\mathcal L = \gamma \boldsymbol{\mathcal J}^2 +(1-\gamma)\left(\mathcal J_z^{(1)}\mathcal J_z^{(2)}+\mathcal J_z^{(2)}\mathcal J_z^{(1)}\right) .
\end{equation}
Our task is to diagonalize the matrix representation of this superoperator. The matrix elements in the basis of spherical tensor products are
\begin{equation}
(\mathcal L)_{m_1m_2,m_1'm_2'}=\frac{\left(T^{(1)}_{1 m_1}\!\!\otimes\! T^{(2)}_{1 m_2}\big| \mathcal L(T^{(1)}_{1 m_1'}\!\!\otimes\! T^{(2)}_{1 m_2'})\right)}
{\big|\big|T^{(1)}_{1 m_1} \!\otimes\! T^{(2)}_{1 m_2}\big|\big|\,\big|\big|T^{(1)}_{1 m_1'}\!\!\otimes\! T^{(2)}_{1 m_2'}\big|\big|} ,
\end{equation}
where  $(A|B)=\mathrm{Tr}( A^\dagger B)$ and $||A||=\sqrt{(A|A)}$. The explicit form of the matrix is given by
\begin{widetext}
\begin{equation}
\mathcal L=2\hspace{.2cm}\begin{blockarray}{cccccccccc}
 \phantom{-}1 \phantom{-}1 &  \phantom{-}1 \phantom{-}0 &  \phantom{-}0 \phantom{-}1 &  \phantom{-}1{-1} & \phantom{-}0\phantom{-}0 & {-1}\phantom{-}1& \phantom{-}0{-1}& {-1}\phantom{-}0&{-1}{-1}&\\[.1cm]
\begin{block}{(ccccccccc)@{\hspace{.4cm}}c}
1+2\gamma &   &   &   &   &   &   &   & &  \phantom{-}1 \phantom{-}1\\
  & 2\gamma  & \gamma  &   &   &   &   &   &  & \phantom{-}1 \phantom{-}0\\
  & \gamma  & 2\gamma  &   &   &   &   &   & & \phantom{-}0 \phantom{-}1\\
  &   &   & -1+2\gamma  & \gamma  &   &   &   & & \phantom{-}1{-1}\\
  &   &   & \gamma  & 2\gamma  & \gamma  &   &   & & \phantom{-}0 \phantom{-}0\\
  &   &   &   & \gamma  & -1+2\gamma  &   &   & &{-1}\phantom{-}1\\
  &   &   &   &   &   & 2\gamma  & \gamma  & & \phantom{-}0{-1}\\
  &   &   &   &   &   & \gamma  & 2\gamma  & &{-1}\phantom{-}0\\
  &   &   &   &   &   &   &   & 1+2\gamma&{-1}{-1}\\
\end{block}
\end{blockarray} \, \, ,
\end{equation}
\end{widetext}
where blank spaces are zero. The matrix has a block-diagonal form and each block can be diagonalized analytically. In particular, spherical tensor products $T^{(1)}_{1 \pm 1}\!\!\otimes\! T^{(2)}_{1 \pm1} = T_{2\pm 2(11)}$ are already eigenoperators of $\mathcal L$ and the triple of unpolarized products $\{ T^{(1)}_{1 0}~{\!\!\otimes\!}~ T^{(2)}_{1 0}, T^{(1)}_{1 1}~{\!\!\otimes\!}~T^{(2)}_{1 {-1}}, T^{(1)}_{1 {-1}}~{\!\!\otimes\!}~ T^{(2)}_{1 1} \}$ form one of the blocks. Because the $X_{\mathrm{corr}}$--states (including the Bell states) are spanned by these five operators, it follows that any state which belongs to this class remains in it for all time when it is evolved with $\mathcal U_{\mathrm{tr}}$.

Below we list the eigenoperators and the corresponding eigenvalues of $\mathcal L$:
\begin{align}
2(1+2\gamma):{}&\; T^{(1)}_{1 \pm 1}\!\!\otimes\! T^{(2)}_{1 \pm1}\\
-2(1-2\gamma):{}&\; -T^{(1)}_{1 1}\!\!\otimes\! T^{(2)}_{1 -1}+T^{(1)}_{1 -1}\!\!\otimes\! T^{(2)}_{1 1} , \\
-1+4\gamma-3\eta:{}&\;T^{(1)}_{1 1}\!\!\otimes\! T^{(2)}_{1 -1}+T^{(1)}_{1 -1}\!\!\otimes\! T^{(2)}_{1 1}+\nonumber\\
&+\frac{1-3\eta}{2\gamma}T^{(1)}_{1 0}\!\!\otimes\! T^{(2)}_{1 0} , \\
-1+4\gamma+3\eta:{}&\;T^{(1)}_{1 1}\!\!\otimes\! T^{(2)}_{1 -1}+T^{(1)}_{1 -1}\!\!\otimes\! T^{(2)}_{1 1}+\nonumber\\
&+\frac{1+3\eta}{2\gamma}T^{(1)}_{1 0}\!\!\otimes\! T^{(2)}_{1 0} , \\
6\gamma:{}&\; T^{(1)}_{1 0}\!\!\otimes\! T^{(2)}_{1 \pm 1}+T^{(1)}_{1 \pm1}\!\!\otimes\! T^{(2)}_{1 0} , \\
2\gamma:{}&\; T^{(1)}_{1 0}\!\!\otimes\! T^{(2)}_{1 \pm1}-T^{(1)}_{1 \pm1}\!\!\otimes\! T^{(2)}_{1 0} ,
\end{align}
where $\eta = \frac 13\sqrt{1+8\gamma^2}$.

\bibliography{refs_quant,refs_entanglement,refs_added_by_YB}

\end{document}